\documentclass[11pt]{article}
\usepackage{amssymb,dsfont,amsmath,amsfonts,hyperref,mdframed,mathrsfs,stmaryrd}
\usepackage[active]{srcltx}
\usepackage{slashed}
\usepackage{color}
\advance \textheight by \topskip
\let\oldsqrt\sqrt
\def\sqrt{\mathpalette\DHLhksqrt}
\def\DHLhksqrt#1#2{%
\setbox0=\hbox{$#1\oldsqrt{#2\,}$}\dimen0=\ht0
\advance\dimen0-0.2\ht0
\setbox2=\hbox{\vrule height\ht0 depth -\dimen0}%
{\box0\lower0.4pt\box2}}

\def\be{\begin{equation}}
\def\ee{\end{equation}}
\def\bea{\begin{eqnarray}}
\def\eea{\end{eqnarray}}

\def\a{\alpha}
\def\b{\beta}
\def\g{\gamma}
\def\d{\delta}
\def\e{\epsilon}
\def\l{\lambda}
\def\s{\sigma}

\def\Real{\mathbb{R}}

\begin{document}

\begin{center}
{\Large\bf
Gravitational and gauge couplings in \\ Chern--Simons fractional spin gravity}\\
\vspace*{15mm}
Nicolas Boulanger\footnote{\texttt{nicolas.boulanger@umons.ac.be}; Associate
researcher of the FNRS, Belgium.}\\
\vspace*{2mm}
{\textit{M\'ecanique et Gravitation}}\\
{\textit{Universit\'e de Mons -- UMONS, 20 Place du Parc, 7000 Mons, Belgique}} \\
{\textit{and}}\\
{\textit{Laboratoire de Math\'ematiques et Physique Th\'eorique}}\\
{\textit{Unit\'e Mixte de Recherche 7350 du CNRS }}\\
{\textit{Universit\'e Fran{\c{c}}ois Rabelais, Parc de Grandmont, 37200 Tours, France}} \\
\vspace*{5mm}
Per Sundell\footnote{\texttt{per.sundell@unab.cl}}\\
\vspace*{2mm}
{\textit{Departamento de Ciencias F\'isicas}}\\
{\textit{Universidad Andres Bello, Republica 220, Santiago, Chile}}\\
\vspace*{5mm}
Mauricio Valenzuela
\footnote{\texttt{valenzuela.u@gmail.com}}\\
\vspace*{2mm}
{\textit{Facultad de Ingenier\'ia y Tecnolog\'ia}}\\
{\textit{Universidad San Sebasti\'an, General Lagos 1163, Valdivia 5110693, Chile}}

\end{center}

\vspace{1cm}

\begin{minipage}{.90\textwidth}

\paragraph{Abstract}
We propose an extension of Vasiliev's supertrace
operation for the enveloping algebra of Wigner's
deformed oscillator algebra to the fractional
spin algebra given in arXiv:1312.5700.
We provide a necessary and sufficient condition 
for the consistency of the supertrace, 
through the existence of a certain ground state projector. 
We build this projector and check its properties to 
the first two orders in the number operator and to all 
orders in the deformation parameter. 
We then find the relation 
between the gravitational and internal gauge couplings
in the resulting unified three-dimensional Chern--Simons
theory for Blencowe--Vasiliev higher
spin gravity coupled to fractional spin fields
and internal gauge potentials.
We also examine the model for integer or half-integer 
fractional spins, where infinite dimensional ideals 
arise and decouple, leaving finite dimensional gauge 
algebras $gl(2\ell+1)$ or $gl(\ell|\ell+1)$ and
various real forms thereof.

\end{minipage}

\renewcommand{\thefootnote}{\arabic{footnote}}

\setcounter{footnote}{0}

\newpage

{\small \tableofcontents }

\numberwithin{equation}{section}

\section{Introduction}

\subsection{General background}

Higher spin gravity \cite{Vasiliev:1990en,Vasiliev:1992av}
was originally aimed as a theory generalizing gravitational
interactions to arbitrary spin, understood to be integer or
half-an-integer.
However, in constantly curved $2+1$-dimensional backgrounds,
one may lift this restriction, as there the isometry algebras admit physical representations
with fractional spins \cite{Bargmann:1946me,Barut:1965}, \emph{i.e.}
spins interpolating between half-integer and integer numbers, which can be carried by particles carrying anyon statistics \cite{Leinaas:1977fm,Wilczek:1982wy};
see \cite{Forte:1990hd} for a review.
Thus, aiming at a complete description of interacting relativistic fields
in three dimensions, it is natural to ask whether fractional spin fields can be
coupled to the gravitational field and internal gauge fields.

In \cite{Boulanger:2013naa}, we proposed an action describing
non-Abelian interactions among fractional spin fields,
tensorial higher spin fields and internal gauge fields
using a flat connection valued in a fractional spin algebra.
The tensorial higher spin gauge fields are packaged into a
Blencowe--Vasiliev \cite{Blencowe:1988gj,Vasiliev:1989re} master one-form $W$
valued in a bosonic higher spin algebra forming a subalgebra
of the fractional spin algebra and whose basis is given in terms of monomials
in Wigner-deformed oscillators.
Likewise, the internal gauge fields make up a bosonic one-form
$U$ valued in a compact real form of an infinite dimensional
matrix algebra, formally isomorphic to an algebra of Fock space
endomorphisms.
As for the fractional spin fields, they are collected into
two master fields $(\psi,\overline\psi)$ that intertwine
the gravitational and internal gauge algebras.
Thus, the model constructed in \cite{Boulanger:2013naa} can
be regarded in two ways: Either as an extension of the Blencowe--Vasiliev
theory \cite{Blencowe:1988gj,Vasiliev:1989re} by internal and
fractional spin fields, or as a formal (real) analytical
continuation of Chern-Simons supergravity in 2+1 dimensions
\cite{Achucarro:1987vz,Achucarro:1989gm} whereby the gravitino fields are
extended into fractional spin fields, thus forcing the
introduction of fields with spin greater than two.

On shell, the consistency of the construction relies on
the associativity of the underlying fractional spin
algebra, which was demonstrated in \cite{Boulanger:2013naa}
using Fock space methods.
Off shell, however, the action requires a trace operation,
whose precise form was conjectured in \cite{Boulanger:2013naa},
and whose explicit construction we shall tend to below.
Our construction is facilitated by extending Vasiliev's
supertrace operation \cite{Vasiliev:1989re} on the higher spin algebra from its
original domain of validity, which is the space of polynomials
in the generators of the deformed oscillator algebra, to
the space of non-polynomial elements that spans
the internal gauge algebra and the intertwiners.
As a concrete application of this result, we compute
the explicit relation between the gravitational and
gauge couplings in the resulting unified model.

\subsection{Problem setting and main results}

The field content of the model can be assembled into a
matrix master field
\be \mathbb A=\left[\begin{array}{cc} W&\psi\\[5pt] \overline\psi&U
\end{array}\right]\ ,\ee
valued in an associative\footnote{The model can
be projected further to a model based on a Lie algebra.
However, thinking of it as a truncation of
a larger theory containing also matter fields,
the notion of an associative gauge algebra becomes
crucial.}
fractional spin algebra ${\cal A}_\pm$,
containing semiclassical bosonic ($-$) or fermionic ($+$)
fractional spin fields.
The action proposed in \cite{Boulanger:2013naa} is of
the standard Chern--Simons format, \emph{viz.}
\be S_\pm[\mathbb A]=\frac{\varkappa}{2\pi}\int_{M_3} {\rm Tr}_{{\cal A}_\pm} \left(\tfrac{1}{2}\, \mathbb A\star d\mathbb A+
\tfrac{1}{3}\, \mathbb A\star \mathbb A \star \mathbb A\right)\ ,\label{action}\ee
where thus the key ingredients are the
associative product $\star$ (including the wedge product)
and a non-degenerate cyclic trace operation ${\rm Tr}_{{\cal A}_\pm}$.
Assuming their salient features, the resulting equations of motion read
\be d\mathbb A+\mathbb A\star \mathbb A=0\ ,\label{eom}\ee
or in components,
\be dW+W\star W+\psi\star \overline\psi=0\ ,\qquad dU+U\star U+\overline \psi\star \psi=0\ ,\label{eom1}\ee
\be d\psi+W\star \psi+\psi\star U=0\ ,\qquad d\overline\psi+U\star \overline\psi+\overline\psi\star W=0\ .\label{eom2}\ee
As found in \cite{Boulanger:2013naa},
the standard bra-ket formalism for Fock space
endomorphisms can be augmented by a set of
fusion rules, reflecting the couplings in Eqs.
\eqref{eom1}--\eqref{eom2}, that suffice for
on-shell consistency.
More precisely, the fusion rules stipulate how to
perform the star products in Eqs. \eqref{eom1}
and \eqref{eom2} and expand the results in the
bases stipulated by the linear terms while
preserving associativity so as to achieve a
Cartan integrable system.

The off-shell formulation, however, requires an algebraic machinery
that facilitates a trace operation that
applies to both higher spin and internal matrix subalgebras.
The standard (unregularized) Fock space trace operation does not
suffice as it does not apply straightforwardly to polynomials
in deformed oscillators\footnote{To our best understanding, it
remains unclear whether a regularized Fock space trace
operation could be used to construct ${\rm Tr}_{{\cal A}_\pm}$.}.
Instead, in \cite{Boulanger:2013naa}, it was proposed
to realize the internal matrix subalgebra of ${\cal A}_\pm$
using real-analytic non-polynomial symbols and obtaining 
${\rm Tr}_{{\cal A}_\pm}$ by extending Vasiliev's supertrace 
operation \cite{Vasiliev:1989re} correspondingly,
beyond its original domain of validity (given by the algebra of
arbitrary polynomials in the deformed oscillators).
To this end, in order to demonstrate the salient
features of the star product and trace operation
on ${\cal A}_\pm$, it suffices to establish the following conditions:\\[5pt]
\noindent i) Finite star products 
in ${\cal A}_\pm$ (from which  
associativity follows);\\[5pt]
\noindent ii) Finite Vasiliev supertraces 
(which together with (i) implies cyclicity).\\[5pt]
In what follows, we shall show that the matrix
subalgebra of ${\cal A}_\pm$ is realized in terms
of confluent hypergeometric functions of the spin operator,
which establishes (ii).
We shall also verify a necessary condition for 
(i) to hold true, namely that the supertrace of
the square of two ground state projectors is
finite, hoping to present a complete proof together
with a construction of a convolution formula for 
the deformed oscillator star product in Weyl order
and a related trace formula in a future work.

Moreover, assuming the consistency of the model,
and focusing on the fermion model (with
internal $k$-parity $\s=-1$), we shall compute the
aforementioned relations between couplings, \emph{viz.}
$$k_{\rm hs}=\frac{\varkappa}{16}(1-\nu^2)(1-\frac\nu 3)\ ,\qquad
k_{\rm int}=\varkappa \ ,$$
between the fractional spin coupling $\varkappa$
defined in \eqref{action}, the higher spin
coupling $k_{\rm hs}$ (related to Newton's constant
$G_N$ as in \eqref{GN}) and the level $k_{\rm int}$ of
the internal gauge theory.

The paper is organised as follows: In Section \ref{sec:Definition},
we define the algebraic structures of our model.
In Section \ref{sec:Construction}, we construct the symbols
of the basis elements of the internal and fractional spin
sectors of the theory, analyse the properties of the
extension of Vasiliev's trace operation and consider
critical limits.
Reality conditions and various projections are studied in
Section \ref{sec:Real},
where we also establish the aforementioned relation between
the gravitational and internal couplings.
Finally, we conclude in Section \ref{sec:Conclusion}.
Our three-dimensional spinor conventions are given in the Appendix.

\section{Definition of the model}\label{sec:Definition}

In this section we detail the basic algebraic structures
going into the model.

\subsection{Deformed oscillator enveloping algebra and supertrace}

The higher spin fields are thus collected in a Blencowe--Vasiliev
master one-form
\be W\in {\cal W}^{++}\otimes {\rm Cliff}(\gamma)\ ,\ee
where $\gamma$ is an idempotent element introduced to
account for the anti-de Sitter translations, and
${\cal W}^{++}$ is an associative algebra given by a certain
non-polynomial extension, to be spelled out in Section
\ref{sec:npolyext}, of the enveloping algebra $Aq(2;\nu)$ \cite{Vasiliev:1989re}
of the deformed oscillator algebra \cite{Wigner:50,Yang:51}\footnote{
As our analysis only relies on the fundamental relations given in \eqref{do1},
it remains valid for any realization of the
deformed oscillators. } (see also \cite{Plyushchay:1994re,Plyushchay:1997ty})
\be [q_\a,q_\b]_\star=2i(1+\nu k)\e_{\a\b}\ ,\qquad
\{k,q_\a\}_\star=0\ ,\qquad k\star k=1\ ,\label{do1}\ee\be
(q_\a)^\dagger=q_\a\ ,\qquad k^\dagger = k\ ,\quad \nu\in\mathbb R\ .\label{do2}\ee
By its definition, the algebra $Aq(2;\nu)$ consists of
arbitrary star polynomials (of finite degree) in $(q_\a,k)$.
In addition to the hermitian conjugation, which acts
as $(f\star g)^\dagger=g^\dagger\star f^\dagger$, this algebra
has a linear anti-involution $\tau$ defined by
\be \tau(f\star g)=\tau(g)\star \tau(f)\ ,\qquad \tau(q_\a,k)=(iq_\a,k)\ .
\label{taumap}\ee
In general, certain algebraic properties of
infinite-dimensional associative algebras,
such as unitarity and indecomposability of
representations, crucially depend on the
choice of basis.
As basis for ${Aq}(2;\nu)$ we choose the
Weyl-ordered elements
\be T_{\a(n)} := q_{\a_1} \cdots q_{\a_n}\equiv q_{(\a_1}\star\cdots\star q_{\a_n)}\ ,\ee
and $T_{\a(n)}\star k$, where the symmetrization has unit strength.
Equivalently, one may use the projected elements
\be \label{PTP}
T^{\sigma,\sigma'}_{\a(n)} := \left[q_{\a_1} \cdots q_{\a_n}\right]^{\s,\s'}
:=\Pi^{\sigma}\star q_{(\a_1}\star\cdots\star q_{\a_n)}\star  \Pi^{\sigma'}
\ , \qquad \Pi^\pm=\frac{1}{2} (1\pm k)\,,
\ee
which are non-vanishing iff $\sigma\sigma'=(-1)^n$.
Correspondingly, we define the projections
\be {Aq}(2;\nu)^{\sigma,\sigma'}
=\Pi^{\sigma}\star{Aq}(2;\nu)\star\Pi^{\sigma'}\ .\quad
\ee
One may thus represent the elements $f\in Aq(2;\nu)$
by polynomials in the semi-classical basis elements
$q_{\a_1} \cdots q_{\a_n}$ and $q_{\a_1} \cdots q_{\a_n}k:=q_{\a_1} \cdots q_{\a_n}\star k$, referred to as their Weyl ordered symbols,
and which we shall denote by $f$ as well, in a
slight abuse of the otherwise more involved notation.
Thus, using this representation, the operator product
amounts to a non-local composition rule for symbols,
which we shall denote by a $\star$ as well.
As far as star product compositions of monomials
are concerned, they can be deduced by iterating
\be q_\a\star T_{\b(n)}= T_{\a\b(n)}+in\e_{\a(\b_1} \left(1+  \frac{n+\frac12(1-(-1)^n)}{n(n+1)}\,\nu k\,\right)\star T_{\b(n-1))}\ ,\ee
or its projected form
\be q_\a\star T^{\sigma,\sigma'}_{\b(n)}= T^{-\sigma,\sigma'}_{\a\b(n)}+in\e_{\a(\b_1} \left(1-  \frac{n+\frac12(1-(-1)^n)}{n(n+1)}\,\nu \sigma\,\right)\star T^{-\sigma,\sigma'}_{\b(n-1))}\ .\ee
It follows that $Aq(2;\nu)$ does not contain any ideal for
\be 
\mbox{Non-critical}\ \nu \notin 2\mathbb Z+1\ , 
\ee
while for ($\ell=0,1,2,\dots$; $\hat \s=\pm1$)
\be \mbox{Critical}\ \nu\equiv (2\ell+1)\hat \s \in2\mathbb Z+1\ee
it contains the ideal
\be Aq'(2;\nu)=\biguplus_{n\geqslant 0} \left[T^{\hat\s,\hat\s}_{\a(2\ell+2n)}
\oplus T^{\hat\s,-\hat\s}_{\a(2\ell+1+2n)}\right]\ ,\ee
giving rise to the finite dimensional coset
\be \frac{Aq(2;\nu)}{Aq'(2;\nu)} \cong gl(2\ell+1)= gl(\ell)^{\hat\s,\hat\s}\oplus
gl(\ell+1)^{-\hat\s,-\hat\s}\inplus\left[ (\ell,\ell+1)^{\hat\s,-\hat\s}\oplus
(\ell+1,\ell)^{-\hat\s,\hat\s}\right]\ ,\label{Aqcoset}\ee
where $(\ell,\ell+1)$ and $(\ell+1,\ell)$ denote the $\ell(\ell+1)$-dimensional
bi-fundamental representations of $gl(\ell)\oplus gl(\ell+1)$, realized
as suitably $\Pi^\pm$-projected odd polynomials.

Turning to Vasiliev's cyclic trace operation on ${Aq}(2;\nu)$,
it is given by \cite{Vasiliev:1989re}
\be {\rm Tr}_{{Aq}(2;\nu)} (\cdot)={\rm STr}_{{Aq}(2;\nu)}(k\star (\cdot))\ ,
\label{tr}\ee
where the graded cyclic supertrace operation ${\rm STr}_{{Aq}(2;\nu)}$
is fixed uniquely by its defining properties
\be {\rm STr}_{{Aq}(2;\nu)}(f\star g)={\rm STr}_{{Aq}(2;\nu)}(g\star k\star f\star k)
\ ,\qquad {\rm STr}_{{Aq}(2;\nu)}(1)=1\ .\label{str1}\ee
Thus, if $f$ has a definite parity, \emph{viz.} $k\star f\star k=(-1)^f f $, then
${\rm STr}_{{Aq}(2;\nu)}(f\star g)=(-1)^f{\rm STr}_{{Aq}(2;\nu)}(g\star f)$.
In the Weyl ordered basis, one has
\be {\rm STr}_{{Aq}(2;\nu)}\, T^{\sigma,\sigma'}_{\a(n)}=\delta_{n,0}\, \delta^{\s,\s'}\,
\frac{(1-\s\nu)}{2}\ .\ee
More compactly, by representing $f$ using its Weyl ordered symbol
$f(q,k)$, one has
\be
\label{STr}{\rm STr}_{{Aq}(2;\nu)} (f)= f(0;-\nu) \ ,\ee
that is, the trace operation maps the Kleinian $k$ to $-\nu$
inside the symbol.
In critical limits, one has
\be {\rm STr}_{{Aq}(2;\nu)} Aq'(2;\nu)=0\ ,\ee
which means that in critical limits the model
is truncated to
\be W\in gl(\ell+\frac12(1+\hat\s))\otimes {\rm Cliff}(\gamma)\ .\ee

\subsection{Non-polynomial extension and fractional spin algebra}\label{sec:npolyext}

In order to describe the fractional spin model,
we extend the enveloping algebra $Aq(2;\nu)$
into an associative algebra module $Mq(2;\nu)$
that contains two dual subspaces as follows:
We first introduce the formal associative extension
$\overline{Aq}(2;\nu)$ of $Aq(2;\nu)$ consisting of
elements $f$ with Weyl ordered symbols given by
power series
\be f(q,k)=\sum_{m=0}^{\infty}\sum_{n=0,1}
f^{\a(m)} q_{\a_a}\cdots q_{\a_n}k^n\ ,\label{weyl}\ee
that we shall assume are traceable using the natural
extension of \eqref{tr}.
In order to specify this extension, we need to
introduce suitable dual basis elements, \emph{cf.}
the addition of points at infinity to a non-compact
manifold.
The dual elements form a dual associative algebra
$Aw(2;\nu)$, which will turn out to consist of Wigner
distributions that have fixed eigenvalues under the
one-sided action of the spin operator belonging
to lowest or highest weight spaces.

To this end, we choose the Lorentz connection to be the
gauge field in $W$ associated with the
$so(1,2)\cong sl(2;\Real)$ algebra generated by
\be J_a=\frac14 \,(\tau_a)^{\a\b} J_{\a\b}\ ,\qquad
J_{\a\b}=\frac12 \, q_{(\a}\star q_{\b)} \star \Pi^+\ ,\label{Ja}\ee
using the conventions given in Appendix \ref{app:A}.
By this embedding of the Lorentz algebra into the
gauge algebra, it follows that $T^{++}_{\a(n)}$
($n=0,2,4,\dots$), and hence the corresponding
gauge fields, transform in the adjoint representation of spin $n/2$, which are
thus integers.
The fields $(\psi,\overline\psi)$, on the other
hand, transform under Lorentz transformations in
representations induced by the separate left and
right star multiplication by $J_{\a\b}$, respectively.
Indeed, by examining the Casimir operator $C_2\equiv j(1-j)$,
one finds that these representations are characterized by
a spin $j=\frac{1+\nu}4$, which contains a fractional part,
not given by an integer or half-an-integer, except
for
\be \mbox{critical $\nu\in 2\mathbb Z+1\ .$}\ee
To specify these representations, we assume that they are equipped
by basis states that diagonalize the spatial spin
generator\footnote{
Other classes of fractional spin models arise if one instead
chooses to diagonalize a boost or a light-like spin
generator.}

\be J_0=\frac12 \, w\star \Pi^+\ ,\qquad w=\frac14 \, (\tau_0)^{\a\b}
q_\a\star q_\b\ .\label{spinop}\ee
Furthermore, for a complete specification, one needs
to specify whether and how this operator is bounded or not.
For definiteness, we shall take
\be {Aw}(2;\nu)=
\bigoplus_{\s,\s';\e;\l,\l'} T^{\s,\s'}_{\l|\e|\l'}\ ,
\label{overlineaq}\ee
to consist of finite linear combinations
of generalized  quasi-projectors $T^{\s,\s'}_{\l|\e|\l'}
\in \overline{Aq}(2;\nu)$ obeying the ``$\star$-genvalue" equation
\be (w-\l)\star  T^{\s,\s'}_{\l|\e|\l'}=0=T^{\s,\s'}_{\l|\e|\l'}\star (w-\l')\ ,\qquad (T^{\s,\s'}_{\l|\e|\l'})^\dagger=T^{\s',\s}_{\l'|\e|\l}\ ,\qquad\e=\pm 1\ ,\ee
and belonging to one-sided representations of $Aq(2;\nu)$
in which $\e w$ is bounded from below\footnote{
The algebra $Aw(2;\nu)$ is a subalgebra of the algebra
$Aw_{\rm ext}(2;\nu)$ spanned
by quasi-projectors $T^{\s,\s'}_{\l|(\e,\e')|\l'}$
belonging to one-sided representations of $Aq(2;\nu)$ in which
the left action of $\e w$ and the right action of $\e'w$
are bounded from below; the space $Aw_{\rm ext}(2;\nu)\setminus
Aw(2;\nu)$ thus consists of quasi-projectors that
connect states in lowest (highest) weight spaces
to highest (lowest) weight spaces. For example, for $\nu=0$
one has $T^{\s,\s}_{\tfrac\e 2|(\e,-\e)|-\tfrac\e 2}=\pi \delta(a^{-\e})$.}.
Thus, accounting for the internal gauge fields as well,
one is led to the basic fractional spin algebra
\footnote{In \cite{Boulanger:2013naa} we used an auxiliary Fock space
${\cal F}$ to define the fractional spin subalgebra ${\cal A}(2;\nu|\mathfrak{o}(2)_{J_0};{\cal F})$
of $Aw(2;\nu)$ obtained by restricting
to the subspace in which $w$ is bounded
from below.}
\be {\cal A}(2;\nu|w)
:=\left[\begin{array}{cc} {\cal W}^{++} &
{\cal I}^{+-}\\ \overline{\cal I}^{-+}&
{\cal U}^{--}\end{array}\right]\ ,\label{fsa1}
\ee
consisting of the spaces
\be {\cal W}^{++}=\Pi^+\star \overline{Aq}(2;\nu)\star \Pi^+\ ,\qquad
{\cal U}^{--}=\Pi^-\star \overline{Aw}(2;\nu)\star \Pi^-\ ,\ee
\be {\cal I}^{+-}=\Pi^+\star \overline{Aw}(2;\nu)\star \Pi^-\ ,\qquad
\overline{\cal I}^{-+}=\Pi^-\star \overline{Aw}(2;\nu)\star \Pi^+\ ,\ee
where $\overline{Aw}(2;\nu)$ is the extension of
of $Aw(2;\nu)$ by infinite-dimensional traceable matrices.
The associative product law of ${\cal A}(2;\nu|w)$
is defined by a fusion rule, which one may think of
as a germ of an underlying topological open string,
that stipulates that: i) the product of an
arbitrary polynomial and a quasi-projector
is always to be expanded in the basis of quasi-projectors;
and ii) the product of two quasi-projectors is to
be expanded in terms of the basis of quasi-projectors
or the basis of Weyl-ordered monomials in accordance
with the sector to which the product belongs\footnote{
The fusion rule does not require that a monomial admits
any expansion in the basis of quasi-projectors.}.
Thus, returning to the abstract module, we define it formally as
\be Mq(2;\nu)={Aq}(2;\nu)\cup
\left(\overline{Aq}(2;\nu)\cap\overline{Aw}(2;\nu)\right)\ ,\ee
which thus contains $Aq(2;\nu)$ and $Aw(2;\nu)$ as two
dual subalgebras.
There is an asymmmetry between these two spaces, as the
construction overlap requires $Aw(2;\nu)$ to be
mapped to $\overline{Aq}(2;\nu)$ while it does not
require any converse map.
In other words, the module, thought of as a manifold, is glued
together via a monomorphism $\rho: Aw(2;\nu)\rightarrow \overline{Aq}(2;\nu)$.
Thus, in order to define the trace operation on $Mq(2;\nu)$,
it suffices to introduce a trace operation ${\rm Tr}_{Aq(2;\nu)}$
on $Aq(2;\nu)$ and show that it extends to $\rho(Aw(2;\nu))$, \emph{viz.}
${\rm Tr}_{Aw(2;\nu)}(f)={\rm Tr}_{\overline{Aq}(2;\nu)}(\rho(f))$, to which
we shall turn next.

\subsection{Discrete generators, trace operation
and Chern--Simons action}

To treat the cases of Grassmann even or odd
fractional spin fields uniformally and to account
for anti-de Sitter translations, we introduce
a fermionic generator $\xi$ and a bosonic
generator $\gamma$ whose non-trivial relations are
\be \xi\star\xi=1\ ,\qquad \gamma\star\gamma=1\ ,\ee
and extend the fractional spin algebra \eqref{fsa1} into
\be {\cal A}_\pm=\left[
{\cal A}(2;\nu|w)\otimes
{\rm Cliff}(\gamma)\otimes {\rm Cliff}(\xi)\right]_\pm\ ,\ee
consisting of Grassmann even elements
\be \mathbb X=\left[\begin{array}{cc}X^{++}& X^{+-}\\[5pt]
\overline X^{-+}& X^{--}\end{array}\right]\ee
that obey the internal parity condition
\be \pi_q \pi_\xi (\mathbb X)=\mathbb C_\pm\star \mathbb X
\star \mathbb C_\pm\ ,\qquad \mathbb C_\pm=\left[\begin{array}{cc}
1& 0\\[5pt]
0& \pm 1\end{array}\right]\ ,\label{ss}\ee
where $\pi_q$ and $\pi_\xi$ are the automorpisms
of the star product algebra that reverse the sign
of $q_\a$ and $\xi$, respectively.
Thus, the elements $X^{++}$ and $X^{--}$ are $\xi$
independent and hence $W$ and $U$ have expansions
in terms of bosonic component fields.
In the case of ${\cal A}_-$, the same holds for
$(X^{+-},\overline X^{-+})$ and $(\psi,\overline\psi)$.
In the case of ${\cal A}_+$, the elements
$(X^{+-},\overline X^{-+})$ are linear in $\xi$
and hence $(\psi,\bar\psi)$ have expansions
in terms of fermionic component fields.
In other words, the semi-classical statistics of
the component fields is correlated with the
internal parity defined by the
$\pi_q$ map, such that the components of
parity even elements are bosonic while those
of parity odd elements are fermionic in ${\cal A}_+$
and bosonic in ${\cal A}_-$.
More explicitly,
\bea \mbox{Fermionic fractional spin fields (${\cal A}_+$):}&&(\psi,\bar \psi)= (\Theta \star \xi,\xi\star \overline\Theta)\ ,\label{fmodel}\\[5pt]
\mbox{Bosonic fractional spin fields (${\cal A}_-$):}&&(\psi,\bar \psi)= (\Sigma, \overline\Sigma)\ ,\label{bmodel}\eea
where thus $\Theta$ and $\Sigma$ have expansions
in terms of bosonic symbols in $\overline{Aw}^{+,-}(2;\nu)$
multiplied by component fields that are fermions and bosons,
respectively.

Turning to the trace operation, we use the fact that
the fermionic Clifford algebra ${\rm Cliff}(\xi)$, which by its definition
consists of Grassmann even elements of the form $X=X_0+X_1\xi$,
where thus $X_0$ is a boson and $X_1$ is a fermion, has the
supertrace operation\footnote{The supertrace operation, which is
intrisically bosonic, induces an intrinsically fermionic trace operation
${\rm Tr}_{{\rm Cliff}(\xi)}(X)={\rm STr}_{\rm Cliff(\xi)}(\xi\star X)=-X_1$,
which does not play any role in the present class of models.}
${\rm STr}_{{\rm Cliff}(\xi)}(X)=X_0$, which thus obeys ${\rm STr}_{{\rm Cliff}(\xi)}(X\star X')=
{\rm STr}_{{\rm Cliff}(\xi)}(X'\star \xi\star X\star \xi)$.
The bosonic Clifford algebra ${\rm Cliff}(\gamma)$, which by its definition
consists of Grassmann even elements of the form $Y=Y_0+Y_1\gamma$,
has a two-parameter family of trace operations, namely
${\rm Tr}^{(y,\tilde y)}_{{\rm Cliff}(\gamma)}(Y)=\frac{y}2 (Y_0+ Y_1)+
\frac{\tilde y}2 (Y_0- Y_1)$, where $y, \tilde y\in\mathbb R$.

As for $\overline{Aq}(2;\nu)$ and $\overline{Aw}(2;\nu)$,
we extend Vasiliev's supertrace operation
${\rm STr}_{Aq(2;\nu)}$ given in \eqref{STr}, which
is valid for arbitrary polynomials \emph{stricto sensu},
by a procedure that is formally reminiscent of the
fusion rule: The operator in the argument of the
trace is first expanded in the Weyl ordered basis
\eqref{weyl} and the resulting symbol is then
evaluated using \eqref{STr}.
As we shall show below, this extension of ${\rm STr}_{Aq(2;\nu)}$,
that we shall denote by ${\rm STr}_{\overline{Aq}(2;\nu)}$,
preserves its salient features, \emph{viz.}
\be {\rm STr}_{\overline{Aq}(2;\nu)}(f\star g)={\rm STr}_{\overline{Aq}(2;\nu)}(g\star k\star f\star k)\ ,\ee
and if ${\rm STr}_{\overline{Aq}(2;\nu)}(f\star g)=0$ for all
$g$ then $f=0$.

Combining the operations introduced so far with
the standard trace operation on ${\rm Mat}_2$, we are
led to equip the extended fractional spin algebra with
the following trace operation:
\bea {\rm Tr}^{(x)}_{{\cal A}_\pm} \mathbb X&=& {\rm Tr}_{\overline{Aq}(2;\nu)}
{\rm Tr}^{(1+x,-1+x)}_{{\rm Cliff}_1(\gamma)} {\rm STr}_{{\rm Cliff}_1(\xi)}
{\rm Tr}_{{\rm Mat_2}} \mathbb X\star \mathbb C_{\mp}\\[5pt]
&=& {\rm Tr}_{\overline{Aq}(2;\nu)}
{\rm Tr}^{(1+x,-1+x)}_{{\rm Cliff}_1(\gamma)} {\rm STr}_{{\rm Cliff}_1(\xi)}(X^{++}\mp X^{--})\\[5pt]
&=&{\rm STr}_{\overline{Aq}(2;\nu)} \left(k\star\gamma\star (X^{++}\mp X^{--})\right)|_{\gamma=x}\\[5pt]
&=&{\rm STr}_{\overline{Aq}(2;\nu)} \left(\gamma\star (X^{++}\pm X^{--})\right)|_{\gamma=x}\ ,\label{trA}\eea
where $x\in \mathbb R$ is a chiral symmetry breaking parameter.
In view of the claimed properties of ${\rm STr}_{\overline{Aq}(2;\nu)}$,
we thus have
\be {\rm Tr}^{(x)}_{{\cal A}_\pm} \mathbb X\star \mathbb X'={\rm Tr}^{(x)}_{{\cal A}_\pm} \mathbb X'\star \mathbb X\ ,\ee
and that if ${\rm Tr}^{(x)}_{{\cal A}_\pm} \mathbb X\star \mathbb X'=0$ for all $\mathbb X$ then $\mathbb X'=0$.
In order to expand the action \eqref{action}, one decomposes
\be \mathbb A=\mathbb A_{(L)}\star \tfrac12 (1+\gamma)+
\mathbb A_{(R)}\star \tfrac12(1-\gamma)\ ,\ee
where $\mathbb A_{(c)}$ ($c=L,R$) are $\gamma$-independent.
The action, which is thus the natural fractional spin
generalization of chirally asymmetric Chern-Simons
(super)gravities \cite{Blagojevic:2003vn,Giacomini:2006dr},
thus takes the form
\bea S^{(x)}_\pm[\mathbb A]&=& \frac{\varkappa}{2\pi}\int_{M_3}
{\rm Tr}^{(x)}_{{\cal A}_\pm}\left[\tfrac{1}{2}\, \mathbb A\star d\mathbb A+
\tfrac{1}{3}\, \mathbb A\star \mathbb A\star \mathbb A\right]\\[5pt]
&=&\frac{1+x}2\; S_\pm[\mathbb A_{(L)}] - \frac{1-x}2\;
S_\pm[\mathbb A_{(R)}]\ ,\eea
where the chiral action ($c=L,R$)
\be S_\pm[\mathbb A_{(c)}] = \frac{\varkappa}{2\pi} \int_{M_3}  \left[
{\cal L}_{\rm CS}(W_{(c)})\pm {\cal L}_{\rm CS}(U_{(c)})
+ \tfrac12\,{\rm STr}_{\overline{Aq}(2;\nu)}\left(\psi_{(c)} \star
  D\bar{\psi}_{(c)}\pm \bar{\psi}_{(c)} \star D \psi_{(c)}\right)\right]\ ,
  \label{action2}\ee
is defined in terms of the Chern--Simons Lagrangian
\be {\cal L}_{\rm CS}(W_{(c)}) =
 {\rm STr}_{\overline{Aq}(2;\nu)}\left[ \tfrac12 \,W_{(c)} \star d W_{(c)}+\tfrac{1}{3} \,W_{(c)}\star W_{(c)}\star W_{(c)}\right]\ ,\ee
\emph{idem} ${\cal L}_{\rm CS}(U_{(c)})$ and the covariant derivatives
\be D\psi_{(c)}=d\psi_{(c)}+W_{(c)}\star \psi_{(c)}+\psi_{(c)}\star U_{(c)}\ ,\qquad D\overline\psi_{(c)}=d\overline\psi_{(c)}+U_{(c)}\star \overline\psi_{(c)}+\overline\psi_{(c)}\star W_{(c)}\ .\ee
%

\section{Construction of quasi-projectors}\label{sec:Construction}

In this section we construct the non-polynomial 
elements, or Wigner distributions, in the fractional 
spin algebra.
We shall show that they form an associative 
and traceable algebra
provided that the square of a certain ground
state quasi-projector is finite (see Eq. \eqref{key}),
which is the main hypothesis underlying
our construction, and that we hope to 
demonstrate fully elsewhere. In the present paper, 
we show its validity to the first 2 orders in an  
expansion variable, and each time, to all orders in 
$\nu\,$.

\subsection{Creation and annihilation operator basis}

To construct the quasi-projectors it is convenient to change from
the Lorentz covariant basis \eqref{do2} to a basis
of $O(2)_{J_0}$ covariant deformed creation
and annihilation operators
\be a^\pm = u^{\pm \a}q_\a\ ,\quad u^{+\a} u^-_\a~=~-\frac{i}2\ ,\quad (u^{\pm}_\a)^\dagger~=~u^\mp_\a\ .\ee
These operators obey
\be [ a^-, a^+]_\star~=~1+\nu k\ ,\qquad \{k,a^\pm\}_\star=0
\ ,\qquad ( a^\pm)^\dagger~=~ a^\mp\ ,\label{dha}\ee
from which follows the contraction rules
\bea a^{\pm}\star \left[(a^{\mp})^m (a^{\pm})^n\right]^{\s,\s'}&=&\left[(a^{\mp})^m (a^{\pm})^{n+1}\right]^{-\s,\s'}\nonumber\\&&
\mp \frac{m}2 \left(1-\frac{m+n+\frac12(1-(-1)^{m+n})}{(m+n)(m+n+1)}\nu \s\right)
\left[(a^{\mp})^{m-1} (a^{\pm})^{n}\right]^{-\s,\s'}\ ,\qquad\quad\label{contraction1}
\\
\left[(a^{\mp})^m (a^{\pm})^n\right]^{\s,\s'} \star a^{\pm}
&=&\left[(a^{\mp})^m (a^{\pm})^{n+1}\right]^{\s, - \s'}\nonumber\\&&
\pm \frac{m}2 \left(1-\frac{m+n+\frac12(1-(-1)^{m+n})}{(m+n)(m+n+1)}\,\nu \s\right)
\left[(a^{\mp})^{m-1} (a^{\pm})^{n}\right]^{\s, - \s'}\ ,\qquad\quad\label{contraction2}
\eea
where
\be \left[(a^+)^m  (a^-)^n\right]^{\s,\s'}~\equiv~ (u^{+\a})^m (u^{-\a})^n T^{\s,\s'}_{\a(m+n)}\ , \ee
using a shorthand notation in which
$(u^{\pm \a})^m= u^{\pm\, \a_1} \cdots u^{\pm\, \a_m}$.
In this basis, the spin operator \eqref{spinop} and the
basic commutation rules involving it and the deformed
oscillators take the form
\begin{eqnarray}
\label{number} w=a^+ a^-=\tfrac{1}{2}\, \{ a^- , a^+\}_\star\ ,\qquad
[w,a^\pm]_\star=\pm a^\pm\ .
\end{eqnarray}
Writing $w^m=(a^+)^m  (a^-)^m$, one also has the useful relations
\be
a^\pm \star \left[w^m\right]^{\s,\s}=\left[a^\pm \left(w^m \mp \frac{m(2m+1-\nu\s)}{2(2m+1)} w^{m-1}\right)\right]^{-\s,\s}\ ,
\label{contraction3}
\ee
and
\bea
w\star \left[w^m\right]^{\s,\s}&=&\left[w^{m+1}+ \lambda^\s_m
w^ {m-1}\right]^{\s,\s}\ ,
\label{wcontraction}\\[5pt]
 a^{\e}\star \left[w^m\right]^{\s,\s}\star a^{-\e}&=&\left[w^{m+1}+ \e \,\mu^\s_{m} w^m+
\tilde\l^\s_{m} w^ {m-1}\right]^{-\s,-\s}\ ,\label{sandwich}\eea
where we have defined
\begin{equation}\label{lambdam}
\lambda^\s_m=-\frac{m^2}4 \frac{(2m+1-\nu\s)(2m-1+\nu\s)}{(2m+1)(2m-1)}
\ ,
\end{equation}
\be \mu^\s_m=-\frac12 (2m+1-\nu\s)\ ,\qquad
\tilde\l^\s_m=\frac{m^2}4 \frac{(2m+1-\nu\s)(2m-1-\nu\s)}{(2m+1)(2m-1)}\ .\ee

\subsection{Non-critical versus critical $\nu$}

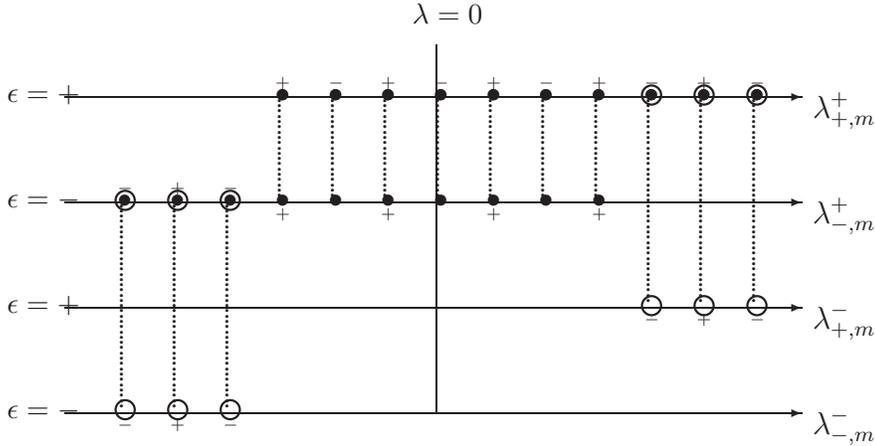
\begin{figure}[ht]
\setlength{\unitlength}{.7cm}
\begin{picture}(10,10)(-3,0)
\multiput(-1.1,-.1)(0,4){2}{$\e=-$}
\multiput(-1.1,1.9)(0,4){2}{$\e=+$}
\multiput(0,2)(0,4){2}{\vector(1,0){14}}
\multiput(0,0)(0,4){2}{\vector(1,0){14}}
\put(7.05,0){\line(0,1){7}}
\put(6.6,7.4){$\l=0$}
\multiput(4,5.9)(1,0){10}{$\bullet$}
\multiput(4,5.9)(1,0){10}{$\bullet$}
\multiput(10.88,5.78)(1,0){3}{\huge$\circ$}
\multiput(10,3.9)(-1,0){10}{$\bullet$}
\multiput(2.88,3.78)(-1,0){3}{\huge$\circ$}
\multiput(10.88,1.78)(1,0){3}{\huge$\circ$}
\multiput(2.88,-.2)(-1,0){3}{\huge$\circ$}
\put(14.2,-.4){\large$\l^-_{-,m}$}
\put(14.2,1.6){\large$\l^-_{+,m}$}
\put(14.2,3.6){\large$\l^+_{-,m}$}
\put(14.2,5.6){\large$\l^+_{+,m}$}
\multiput(1,0)(0,.1){40}{$\cdot$}
\multiput(2,0)(0,.1){40}{$\cdot$}
\multiput(3,0)(0,.1){40}{$\cdot$}
\multiput(11,2)(0,.1){40}{$\cdot$}
\multiput(12,2)(0,.1){40}{$\cdot$}
\multiput(13,2)(0,.1){40}{$\cdot$}
\multiput(4,6.2)(2,0){5}{\tiny$+$}
\multiput(5,6.2)(2,0){5}{\tiny$-$}
\multiput(10,3.7)(-2,0){4}{\tiny$+$}
\multiput(3,4.2)(-2,0){2}{\tiny$-$}
\multiput(2,4.2)(2,0){1}{\tiny$+$}
\multiput(11,1.7)(2,0){2}{\tiny$-$}
\multiput(12,1.7)(2,0){1}{\tiny$+$}
\multiput(3,-.3)(-2,0){2}{\tiny$-$}
\multiput(2,-.3)(2,0){1}{\tiny$+$}
\multiput(4,4)(0,.1){20}{$\cdot$}
\multiput(5,4)(0,.1){20}{$\cdot$}
\multiput(6,4)(0,.1){20}{$\cdot$}
\multiput(7,4)(0,.1){20}{$\cdot$}
\multiput(8,4)(0,.1){20}{$\cdot$}
\multiput(9,4)(0,.1){20}{$\cdot$}
\multiput(10,4)(0,.1){20}{$\cdot$}
\end{picture}
\caption{\small Distribution of the eigenvalues $\l^\s_{\e,m}$ of $w$
for $\ell=3$ and $\hat \s=-1$.
The dashed vertical lines
indicate relations between generalized quasi-projectors.
In the finite dimensional sector, these elements span two-dimensional
solution spaces.
In the infinite dimensional sector, these are proportional
modulo the rescalings given in \eqref{scale1} and \eqref{scale2}. }
\label{fig1}
\end{figure}

From \eqref{wcontraction} and \eqref{lambdam} it follows that if
\be \nu\notin 2\mathbb Z+1\qquad \mbox{(non-critical $\nu$)}\ ,\ee
then all eigenvalues of $w$ are non-degenerate and all quasi-projectors
are descendants of the ground state quasi-projectors
\be T^\s_\e\equiv T^{\s,\s}_{\l^\s_{\e,0}|\e|\l^\s_{\e,0}}\ ,\ee
obeying
\be a^{-\e}\star T^\s_\e=0\ ,\qquad \l^\s_{\e,0}=\frac{\e(1+\nu\s)}2\ ,\ee
which makes $T^\s_\e$ into a lowest (highest) weight state for
$\e=+$ ($\e=-$).
The resulting spectrum of generalized quasi-projectors
and corresponding eigenvalues is given by
\be T^{\s_m,\s_n}_{\l^\s_{\e,m}|\e|\l^\s_{\e,n}} :=
(a^{\e})^{\star m}\star T^\s_\e\star (a^{-\e})^{\star n}\ ,\qquad
m,n\in\{0,1,2,\dots\}\ ,\ee
\label{quasiT}\be
\l^\s_{\e,m}=\e\left(m+\frac12(1+\nu\s)\right)\ ,\qquad \s_m=(-1)^m\s\ .
\label{spectrum}\ee
However, when
\be \nu\in 2\mathbb Z+1\qquad \mbox{(critical $\nu$)}\ ,\ee
then the spectrum degenerates, as illustrated in Figure \ref{fig1},
and singular quasi-projectors arise, leading to an indecomposable
structure to be examined below once
we have completed the definition of the model.

\subsection{Projectors in non-critical case}

For non-critical $\nu$, we define the ground state quasi-projectors
\begin{equation}
T^{\sigma}_\epsilon=\sum_{m=0}^\infty f_m w^m\star\Pi^\sigma\ ,\qquad f_0 = 1\ ,
\end{equation}
obeying
\be a^{-\epsilon}\star T^{\sigma}_\epsilon=0=T^{\sigma}_\epsilon\star a^{\e}\ ,\qquad
{\rm STr}_{\overline{Aq}(2;\nu)} T^{\sigma}_\epsilon= \frac12(1-\nu\sigma)\ .\label{condT}\ee
Using \eqref{contraction3}, one finds a recursive relation for $f_m$ with
a unique solution with $f_0=1$, given by
\begin{equation}
f_m= \frac{(-2\epsilon)^m}{m!} \frac{\left(\frac32\right)_m}{\left(\frac{3-\nu\sigma}2\right)_m}\ ,
\end{equation}
where the Pochhammer symbol $(a)_n$ is given by $1$ if $n=0$ 
and by $a(a+1)\cdots (a+n-1)$ if $n=1,2,\dots$
Hence, using the definition of the confluent hypergeometric function, \emph{viz.}
\be {}_1 F_1(a;b;z)=\sum_{n\geqslant 0} \frac{(a)_n}{(b)_n}\frac{z^n}{n!}\ ,\ee
we have
\begin{equation}
T^{\sigma}_\epsilon = {}_1 F_1\left(\frac32;\frac{3-\sigma\nu}2;-2\epsilon w\right)\star \Pi^\sigma\ ,
\end{equation}
which obeys $(w-\l^\s_{\e,0})\star T^\s_\e=0$ by virtue of 
\eqref{wcontraction}.

The rest of our analysis will be based on the assumption that  
$T^{\sigma}_\epsilon\star
T^{\sigma}_\epsilon$ is finite and non-vanishing, which is 
equivalent to that 
\be T^{\sigma}_\epsilon\star
T^{\sigma}_\epsilon = ({\cal N}^\sigma_\epsilon)^{-1}\, 
T^{\sigma}_\epsilon\ ,\qquad
{\cal N}^\sigma_\epsilon \; \in \;\; ]0,\infty[\ ,
\label{key}\ee
in view of the uniqueness of the solution to \eqref{condT}.
The normalized ground state projector is then defined as follows: %
\be 
P^{\sigma,\sigma}_{\lambda^\sigma_{\epsilon,0}|\epsilon|\lambda^\sigma_{\epsilon,0}}\equiv P^\sigma_\epsilon 
= {\cal N}^\sigma_\epsilon \,T^{\sigma}_\epsilon\ ,\qquad P^\sigma_\epsilon\star P^\sigma_\epsilon 
=P^\sigma_\epsilon\ .
\label{Proj0}\ee
In this section, we will prove that 
\eqref{key} holds true to zeroth and first order in the 
variable $w\,$, and this, to all orders in $\nu\,$. 
That the equation \eqref{key} is true to all orders 
in $w\,$ can be proven by a direct, though tedious 
analysis that we will  present elsewhere.

Let us first prove \eqref{key} to zeroth order in $w\,$.  
The normalization can be obtained from
\begin{equation}
({\cal N}^\sigma_\epsilon)^2 {\rm STr}_{\overline{Aq}(2;\nu)} \,(T^{\sigma}_\epsilon\star
T^{\sigma}_\epsilon) ={\cal N}^\sigma_\epsilon {\rm STr}_{\overline{Aq}(2;\nu)}\,T^{\sigma}_\epsilon\ ,
\label{normalizationcondn}\end{equation}
using
\begin{equation}
{\rm  STr}_{\overline{Aq}(2;\nu)}T^{\sigma}_\epsilon= {\rm  STr}_{\overline{Aq}(2;\nu)}\Pi^\sigma =\frac{1}{2} (1-\nu \sigma)\ ,
\end{equation}
and
\begin{equation}
{\rm STr}_{\overline{Aq}(2;\nu)}(T^{\sigma}_\epsilon\star T^{\sigma}_\epsilon) =
\sum_{m,n} f_m f_n {\rm STr}_{Aq(2;\nu)} (w^m \star w^n\star \Pi^\sigma) \ ,
\end{equation}
where
\begin{equation}
{\rm STr}_{Aq(2;\nu)} (w^m\star w^n\star \Pi^\sigma) = \delta_{m,n} {\rm STr}_{Aq(2;\nu)} (w^m\star w^m\star \Pi^\sigma) \ ,
\end{equation}
from which it follows that
\begin{equation}
{\rm STr}_{Aq(2;\nu)} (w^m\star w^m\star \Pi^\sigma) ={\rm STr}_{Aq(2;\nu)} ( \underbrace{w\star\cdots\star w}_{\mbox{$m$ times}} \star w^m
\star\Pi^\sigma) =
\frac12(1-\sigma\nu) \prod_{n=1}^m \lambda_n^\sigma\ ,
\end{equation}
where we have used \eqref{wcontraction} and \eqref{lambdam}.
Hence, 
\begin{eqnarray}
({\cal N}^\sigma_\epsilon)^{-1}&=&\sum_{m=0}^\infty (f_m)^2 \prod_{n=1}^m \lambda_n^\sigma\label{series}\\[5pt]
&=& 2\,{}_2F_1(\frac{1+\nu\sigma}2,2;\frac{3-\nu\sigma}2;-1)-
{}_2F_1(\frac{1+\nu\sigma}2,1;\frac{3-\nu\sigma}2;-1)
\\[5pt]&=&2\,{}_2F_1(\frac{3+\nu\sigma}2,2;\frac{3-\nu\sigma}2;-1)\\[5pt]
&=&\frac12(1-\nu\sigma)\ ,\label{STrcalc}
\end{eqnarray}
where we have used Gauss' contiguous relations between hypergeometric functions followed by Kummer's evaluation formula.

In order to prove \eqref{key} to the first order in $w\,$, 
we can use the structure constants of the lone-star 
product \cite{Pope:1989sr} (see also \cite{Fradkin:1990qk}) 
in the form\footnote{The result can also be obtained by brute force 
\cite{Korybut:2014jza,ThomasNic}.}
\begin{equation}
  w^m \star w^n = \sum_{p=0}^{2m} C_p^{(m,n)} w^{m+n-p}
\end{equation}
where
\begin{equation}
  C_p^{(m,n)} = \left(\frac{1}2 \right)^p \frac{1}{p!} \, {\cal N}_p^{m,n}
  \phi_p^{(m,n)}(\nu) \ ,
\end{equation}
\begin{equation}
  {\cal N}_p^{m,n} = \sum_{r=0}^p (-1)^r \binom{p}{r} [m]_r [m]_{p-r} [n]_r
    [n]_{p-r} \ ,
\end{equation}
and
\begin{equation}
  \phi_p^{(m,n)}(\nu) = {_4}F_3 \left[
    \begin{aligned}
      \frac{\nu}2, \ \ \, 1-\frac{\nu}2, \ \ \, -\frac{p}2, \ \ \,
      -\frac{p-1}2 \ \, \\ \frac{1-2m}2, \frac{1-2n}2, n+m-p+\frac{3}2
    \end{aligned}
    ; 1 \right] \ .
\end{equation}
Defining the operator 
$\Delta(\cdot):=\left.\frac{d(\cdot)}{dw}\right\vert_{w=0}\,$, 
to first order in $w$ we have 
\begin{equation}
\Delta (T_\epsilon^\sigma \star T_\epsilon^\sigma)  
  =  2\,\sum_{m=0}^\infty f_{m+1} f_m C_{2m}^{(m,m+1)}\ , 
\end{equation}
giving, after some algebra, 
\begin{eqnarray}
  \Delta ( T_\epsilon^\sigma \star T_\epsilon^\sigma)  
 & = & 2\, \sum_{m=0}^\infty \frac{(-2\epsilon)^{m+1}
    (\tfrac{3}2)_{m+1}}{(m+1)!  (\tfrac{3-\sigma \nu}2)_{m+1}}
  \frac{(-2\epsilon)^m (\tfrac{3}2)_m}{m!  (\tfrac{3-\sigma \nu}2)_m}
  \left( \frac{1}2 \right)^{2m} (-1)^m (2)_m^2
  \frac{(\tfrac{5-\nu}2)_m (\tfrac{3+\nu}2)_m}{(\tfrac{5}2)_m
    (\tfrac{3}2)_m} 
    \nonumber \\ 
    & = & 2\,
  \frac{(-2\epsilon) \, \tfrac{3}2}{\tfrac{3-\sigma \nu}2}
  \sum_{m=0}^\infty \frac{(2)_m (\tfrac{5-\nu}2)_m
    (\tfrac{3+\nu}2)_m}{(\tfrac{5-\sigma \nu}2)_m m! (\tfrac{3-\sigma
      \nu}2)_m} (-1)^m \ .
\end{eqnarray}
Taking  $\sigma = +1\,$ for the sake of definiteness
and without loss of generality (as we can always revert 
the sign of $\nu$ in the final result),
we obtain
\begin{eqnarray}
  \Delta (T_\epsilon^+ \star T_\epsilon^+ ) & = & 
  2\, f_1 \, \sum_{m=0}^\infty \frac{(2)_m
    (\tfrac{3+\nu}2)_m}{m! (\tfrac{3-\nu}2)_m} (-1)^m 
    = 2\, f_1 \,
  \ {_2}F_1 \left[\tfrac{3+\nu}2, 2 ; \tfrac{3-\nu}2 ; -1 
  \right]\ , 
  \nonumber \\ 
  \Longrightarrow\qquad  \Delta(T_\epsilon^+ \star T_\epsilon^+) 
  & = & \frac{1-\nu}{2} \, f_1 \, \ ,
      \label{3.40}
\end{eqnarray}
where we used Kummer's theorem to evaluate the 
hypergeometric function.
We have thus proven the correctness of \eqref{key} to 
zeroth and first orders in $w\,$, and each time,  
to all orders in~$\nu\,$.

Thus, for non-critical $\nu$ and under the assumption made 
on
\eqref{key}, the ground state projectors are given by
\be P^{\sigma}_\e = \frac{2}{1-\nu\sigma}
\;{}_1 F_1\left(\frac32;\frac{3-\sigma\nu}2;-2\epsilon w\right)\star \Pi^\sigma\ .\ee
Their supertraces are given by\footnote{The $\nu$-independence
of \eqref{STrP0} (to be compared with the remark below Eq. (2.41) in \cite{Boulanger:2013naa}) is in accordance with realizing 
$(a^\pm,k)$ using undeformed oscillators $b^\pm$ obeying 
$[b^-,b^+]_\star=1$ \cite{Plyushchay:1997ty,Boulanger:2013naa}.
The vacuum projectors $P^\s_\e$ can then be represented 
in $\overline{Aq}(2;0)$ by $2\exp(-2\e b^+b^-) \star \Pi^\s$
with ${\rm STr}_{\overline{Aq}(2;0)} \left(2\exp(-2\e b^+b^-) \star \Pi^\s\right)=
1$.}
\begin{equation}
{\rm STr}_{\overline{Aq}(2;\nu)}\,P^{\sigma,\sigma}_{\e\l_0|\e\l_0} =1\ .
\label{STrP0}
\end{equation}
From these ground state projectors descend matrices of generalized projectors
\be \left(P^{\s_m,\s_n}_{\e}\right)_{\l^\s_{\e,m}}{}^{\l^\s_{\e,n}}= 
\frac{2}{1-\nu\s}\; {\cal C}^{\s}_{\e,m} \,
{\cal N}^\s_{\e,m}\, {\cal N}^{\s}_{\e,n}\,
T^{\s_m,\s_n}_{\l^\s_{\e,m}|\e|\l^\s_{\e,n}}\ ,
\label{Proj}\ee
where the normalization and conjugation coefficients are
given by
\be
{\cal N}^{\s}_{\e,m}=|\eta^\s_{\e,m}|^{-\tfrac12}\ ,\qquad
{\cal C}^{\s}_{\e,m}=\frac{\eta^\s_{\e,m}}{|\eta^\s_{\e,m}|}\ ,\label{calN}\ee
respectively, where the coefficients 
\be (a^{-\e})^{\star m} \star (a^{\e})^{\star m}\star T^\s_\e
=\eta^\s_{\e,m} T^\s_\e\ .\label{eta}\ee
More explicitly, we have $\eta^\s_{\e,0}=1$ and
\be \eta^\s_{\e,m}=\e^m \prod_{n=1}^m \left(n+\frac12(1-(-1)^n)\s\nu\right)\qquad \mbox{for $m=1,2,\dots$}\ .\label{etasigmaepsm}\ee
The usage of the conjugation coefficients in \eqref{Proj} implies the reality condition
\be \left(\left(P^{\s,\s'}_{\e}\right)_{\l}{}^{\l'}\right)^\dagger
=C\star \left(P^{\s',\s}_{\e}\right)_{\l'}{}^{\l}\star C\ ,
\label{Pdagger}\ee
where the conjugation matrix \cite{Plyushchay:1997ty,Boulanger:2013naa}
\be C=\sum_{m,\e,\s} {\cal C}^{\s}_{{\e,m}}
\left(P^{\s_m,\s_m}_{\e}\right)_{\l^\s_{\e,m}}{}^{\l^\s_{\e,m}}
\ .\label{cmatrix}\ee
Turning to the supertrace of the generalized projectors, they are given by 
\bea {\rm STr}_{\overline{Aq}(2;\nu)}\,\left[P^{\sigma_m,\sigma_n}_{\e}\right]
_{\l^\s_{\e,m}}{}^{\l^{\s}_{\e,n}}&=&
\frac{2\,{\cal C}^{\s}_{\e,m} \,{\cal N}^\s_{\e,m}\, {\cal N}^{\s}_{\e,n}}{1-\nu\s}\;
{\rm STr}_{\overline{Aq}(2;\nu)}\left[
(a^\e)^{\star m}\star T^{\s}_{\e}\star (a^{-\e})^{\star m}\right]\nonumber\\[5pt]
&=&  \frac{2\,{\cal C}^{\s}_{\e,m} \,{\cal N}^\s_{\e,m}\, {\cal N}^{\s}_{\e,n}}{1-\nu\s}\; 
{\rm STr}_{\overline{Aq}(2;\nu)}\left[
(a^\e)^{\star m}\star \sum_{n=0}^m f_n w^n \star (a^{-\e})^{\star m}\right]\nonumber\\[5pt]
&=& \frac{2(-1)^m\,{\cal C}^{\s}_{\e,m} \,{\cal N}^\s_{\e,m}\, {\cal N}^{\s}_{\e,n}}{1-\nu\s}\;  
{\rm STr}_{\overline{Aq}(2;\nu)}\left[
 (a^{-\e})^{\star m}\star(a^\e)^{\star m}\star \sum_{n=0}^m f_n w^n\right]\nonumber \\[5pt]
&=&\frac{2(-1)^m\,{\cal C}^{\s}_{\e,m} \,{\cal N}^\s_{\e,m}\, {\cal N}^{\s}_{\e,n}}{1-\nu\s}\;  
{\rm STr}_{\overline{Aq}(2;\nu)} \left[
(a^{-\e})^{\star m}\star (a^\e)^{\star m}\star T^{\s}_{\e}\right]\ ,\qquad\qquad
\eea
where we have allowed ourselves to exchange the order of sums and
supertraces and used 
\be {\rm STr}_{\overline{Aq}(2;\nu)} \left(
(a^\e)^{\star m}\star w^n \star (a^{-\e})^{\star m}\right)=0= {\rm STr}_{\overline{Aq}(2;\nu)} \left(
(a^{-\e})^{\star m}\star (a^\e)^{\star m}\star w^n \right)\qquad n>m\ ,\ee
and the graded cyclic property of the
supertrace operation in the algebra of polynomials.
Thus, it follows from \eqref{calN} and \eqref{eta} that 
\be {\rm STr}_{\overline{Aq}(2;\nu)}\,\left(P^{\sigma_m,\sigma_n}_{\e}\right)
_{\l^\s_{\e,m}}{}^{\l^{\s}_{\e,n}}=(-1)^m\delta_{mn}\ ,
\label{STrPmn}\ee
as one can indeed verify explicitly by making repeated
use of \eqref{sandwich}, \eqref{calN} and \eqref{etasigmaepsm}.
Finally, under the assumption that \eqref{key} holds true,
it follows from \eqref{Proj} that
\be
\left(P^{\s,\s'}_{\e}\right)_{\l}{}^{\l'}\star
\left(P^{\s'',\s'''}_{\e}\right)_{\l''}{}^{\l'''}
=\delta^{\s'\s''}\delta_{\l''}^{\l'}
\left(P^{\s,\s'''}_{\e}\right)_{\l}{}^{\l'''}\ ,\ee
which in its turn implies the graded cyclic property of
the supertrace operation.

In summary so far, the salient features of the fractional
spin algebra, namely its associativity and traceability using
the extension of Vasiliev's supertrace, thus hinge on the finiteness
of the normalization coefficient ${\cal N}^\s_\e$ in \eqref{key},
which we hope to demonstrate in a forthcoming work by using
a convolution formula for the star product in $Aq(2;\nu)$.
Since we have proven the correctness of \eqref{key}
to the first 2 orders in $w\,$, we believe it holds 
true to all orders.\footnote{Using the basic star product formula to evaluate the 
left-hand side of \eqref{key}, we expect the coefficient of $w^m$
to be a sum of ${}_p F_q(\cdots;\cdots;z)$ functions evaluated at $z=-1$.} 

\subsection{Quasi-projectors for critical $\nu=\pm3,\pm5,\dots$}

Turning to critical $\nu$, we first consider the
case
\be \nu=\hat{\s}(2\ell+1)\ ,\qquad \ell~=~1,2,3,\dots\ ,\qquad \hat\s=\pm 1\ ,\ee
leaving the hyper-critical case $\nu=\pm1$ to the end.
Letting $\widehat T^\s_\e$ denote the corresponding
critical ground state quasi-projectors in $\overline{Aq}(2;\nu)$,
\emph{i.e.}
\be a^{-\e}\star \widehat T^\s_\e=0\ ,\qquad (w-\l^\s_{\e,0})\star \widehat T^\s_\e=0\ ,\qquad \l^{-\hat\s}_{\e,0}=\e\ell\ ,\qquad \l^{\hat\s}_{\e,0}=\e(\ell+1)\ ,\ee
we can take
\be \widehat T^{-\hat\s}_\e= \lim_{\nu\rightarrow(2\ell+1)\hat\s}
\,{}_1 F_1\left(\frac32;\frac{3+\hat\sigma\nu}2;-2\epsilon w\right)\star
\Pi^{-\hat\sigma}=
{}_1 F_1\left(\frac32;\ell+2;-2\epsilon w\right)\star
\Pi^{-\hat\sigma}\ ,\ee
which is a non-singular quasi-projector leading to
the projector
\be \widehat P^{-\hat{\s}}_\e=\frac1{\ell+1} {}_1 F_1\left(\frac32;\ell+2;-2\epsilon w\right)\star
\Pi^{-\hat\sigma}\ .\ee
On the other hand, from
\be \frac{\nu-(2\ell+1)\hat\s}{\left(\frac{3-\hat\sigma\nu}2\right)_n} ~\rightarrow ~\left\{\begin{array}{ll}0& \mbox{for $n\leqslant\ell-1$}\ ,\\[5pt]
\frac{2(-1)^\ell\hat\s}{ (\ell-1)! (n-l)!}&\mbox{for $n\geqslant \ell$}\ ,\end{array}\right.\ee
and
\be \left(\frac{3}2\right)_n~=~\left(\frac{3}2\right)_\ell\left(\frac{3}2+\ell\right)_{n-\ell}\ ,\quad n\geqslant \ell\ ,\ee
it follows that
\be \label{scale1}
\lim_{\nu\rightarrow(2\ell+1)\hat\s}
\left(\nu-(2\ell+1)\hat\s\right) \,{}_1 F_1\left(\frac32;\frac{3-\hat\sigma\nu}2;-2\epsilon w\right)\star
\Pi^{\hat\sigma}= \frac{2^{\ell+1}\e^\ell\hat \s  \left(\frac{3}2\right)_\ell}{\ell! (\ell-1)!} \,\widehat T^{\hat\s}_\e\ ,
\ee
where we have defined
\be\widehat T^{\hat\s}_\e=\left[w^\ell {}_1 F_1(\ell+\frac32;\ell+1;-2\e  w)\right]^{\hat\s,\hat\s}\ .\ee
This element obeys
\be a^{-\e}\star  \widehat T^{\hat\sigma}_{\e}=0\ ,\qquad (w-\e(\ell+1) ) \star \widehat T^{\hat\sigma}_{\e}=0\ ,\ee
which can indeed be checked using \eqref{wcontraction}.
Moreover, by taking the limit of the normalization
condition \eqref{normalizationcondn}, under the
assumption that $\widehat T^{\hat\s}_\e$ is supertraceable,
one finds
\be \widehat T^{\hat\sigma}_{\e}\star\widehat T^{\hat\sigma}_{\e}=0\ , \ee
that we refer to as $\widehat T^{\hat\sigma}_{\e}$
being a singular quasi-projector, which can hence not be
normalized to form a projector.
To analyze descendants, we first use \eqref{wcontraction}
and \eqref{lambdam} to show that if
\be \widehat T^\s_{\l;p}=\sum_{m=p}^\infty f^\s_m w^m\star \Pi^\s\ ,\qquad
(w-\l)\star \widehat T^\s_{\l;p}=0\ ,\qquad f^\s_p=1\ ,\ee
then one has one of the following cases:
\bea (p,\s)&=&(0,\hat\s)\ ,\qquad \l\in\left\{\pm (\ell-1),\pm(\ell-3),\cdots,\pm \tfrac12(1+(-1)^\ell)\right\}\ ,\label{case1}\\[5pt]
(p,\s)&=&(0,-\hat \s)\ ,\qquad \l\in\left\{\pm \ell,\pm(\ell-2),\cdots,\pm \tfrac12(1-(-1)^\ell)\right\}\ ,\label{case2}\\[5pt]
(p,\s)&=&(\ell,\hat\s)\quad\mbox{or}\quad (\ell+1,-\hat\s)\qquad
\mbox{(no condition on $\l$)}\ ,\label{casesp}\eea
and $\widehat T^{\hat\s}_{\l;\ell}$ and $\widehat T^{-\hat\s}_{\l;\ell+1}$ are unique
while $\widehat T^\s_{\l;0}$ belongs to a two-dimensional solution space.
We note that the first case in \eqref{casesp} corresponds
to the singular ground state quasi-projector, \emph{viz.}
\be \widehat T^{\hat\s}_{\e(\ell+1);\ell}= \widehat T^{\hat\s}_\e\ .\ee
The quantization of $\l$ in Eqs. \eqref{case1} and \eqref{case2}
follows from the fact that $\{f^{\hat\s}_m\}_{m=0}^{\ell-1}$
and $\{f^{-\hat\s}_m\}_{m=0}^{\ell}$ must obey the homogenous
equation systems
\bea f^{\hat\s}_{m-1}-\l f^{\hat\s}_m-\l_{m+1}^{\hat\s} f_{m+1}&=&0\ ,\qquad \l_{m}^{\hat\s}=
-\frac{m^2(m^2-\ell^2)}{4m^2-1}\equiv\l_m(\ell)\ ,\\[5pt]
 f^{-\hat\s}_{m-1}-\l f^{-\hat\s}_m-\l_{m+1}^{-\hat\s} f_{m+1}&=&0\ ,\qquad
\l_{m}^{ -\hat\s}=\l_m(\ell+1)\ ,\eea
which have rank $r=\ell+\frac12(1-\hat\s)$,
leading to the characteristic equations
\be {\rm det}\left[\begin{array}{cccccccccc}
                                     \l&-\l_1(r)&0&\cdots&&&&&&\\
                                    -1&\l&-\l_2(r)&0&\cdots&&&&&\\
                                    0&-1&\l&-\l_3(r)&0&\cdots&&&&\\
                                    \vdots&0&-1&\ddots&\ddots&\ddots&&&&\\
                                    & \vdots& 0&\ddots&&&&&&\\
                                    &&\vdots&\ddots&&&&\ddots&\ddots&\\
                                    &&&&&&&\ddots&\l&-\l_{r-1}(r)\\
                                    &&&&&&&&-1&\l
                                    \end{array}\right]=0\ .\ee
Thus, descending from the non-singular quasi-projector
one encounters an $(2\ell+1)$-plet of quasi-projectors
\be \widehat T^{-\hat\s(-1)^m,-\hat\s(-1)^m}_{\e (m-\ell)|\e|\e(m-\ell)}=
(a^{+\e})^{\star m}\star \widehat T^{-\hat\s}_\e \star (a^{-\e})^{\star m}\ ,\qquad m=0,1,\dots,2\ell\ ,\ee
all of which have finite supertraces, that is, the element
$\widehat T^{-\hat\s_m,-\hat\s_m}_{\e (m-\ell)|\e|\e(m-\ell)}$
contains a non-trivial component along 
$\widehat T^{-\hat\s_m}_{\e(m-\ell),0}$.
One also has
\be \tau(\widehat T^{-\hat\s_m,-\hat\s_m}_{\e (m-\ell)|\e|\e(m-\ell)})
=\widehat T^{-\hat\s_m,-\hat\s_m}_{-\e (m-\ell)|-\e|-\e(m-\ell)}\ .\ee
Descending one step beyond $\widehat T^{-\hat\s,-\hat\s}_{\e \ell|\e|\e\ell}$
one reaches the $\widehat T^{\hat\s,\hat\s}_{\e(\ell+1)|\e|
\e(\ell+1)}$ which is proportional to
the singular projector $\widehat T^{\hat\s}_\e$,
\emph{i.e.} there exists a non-trivial coefficient ${\cal C}^{\hat\s}_{\ell}$
such that
\be \widehat T^{\hat\s,\hat\s}_{\e(\ell+1)|\e|
\e(\ell+1)}= {\cal C}^{\hat\s}_{\ell}\widehat T^{\hat\s}_\e\ ,\label{scale2}\ee
which is indeed compatible with the cyclicity of the supertrace.
In summary, we have verified necessary conditions for the
supertraceability of the algebra $Aw(2;\nu)$
in the critical cases $\nu=\pm3,\pm5,\dots$.
The resulting indecomposable structure, illustrated
in Figures \ref{fig2} and \ref{fig3}, is given by
\be Aw(2;\nu)=\frac{Aw(2;\nu)}{Aw'(2;\nu)}\niplus Aw'(2;\nu)\ ,\ee
where the ideal
\be Aw'(2;\nu)=\bigoplus_{\e} \bigoplus_{m,n=0}^\infty \left[\widehat T^{(-1)^m\hat\s,(-1)^n
\hat\s}_{\l^{\hat\s}_{\e,m}|\e|\l^{\hat\s}_{\e,n}}
\oplus \widehat T^{(-1)^m\hat\s,(-1)^n
\hat\s}_{\l^{-\hat\s}_{\e,m+2\ell+1}|\e|\l^{-\hat\s}_{\e,n+2\ell+1}} \right]\ ,\ee
and hence the coset
\bea \frac{Aw(2;\nu)}{Aw'(2;\nu)}&=&\bigoplus_{\e} \bigoplus_{m,n=0}^{2\ell}
\widehat T^{-(-1)^m\hat\s,-(-1)^n
\hat\s}_{\l^{-\hat\s}_{\e,m}|\e|\l^{-\hat\s}_{\e,n}}\\[5pt]&\cong& \bigoplus_{\e} gl(2\ell+1)_\e\\[5pt]
&=&
\bigoplus_{\e}\left(gl(\ell)^{\hat\s,\hat\s}\oplus
gl(\ell+1)^{-\hat\s,-\hat\s}\inplus\left[ (\ell,\ell+1)^{\hat\s,-\hat\s}\oplus
(\ell+1,\ell)^{-\hat\s,\hat\s}\right]\right)_\e\ ,\label{Awcoset}\eea
which we identify as two copies of the coset \eqref{Aqcoset} arising in $Aq(2;\nu)$ in the critical limit.
It is worth mentioning that the appearing of ideal subalgebras in the polynomial basis \eqref{weyl}
for critical $\nu$ was already noticed in \cite{Vasiliev:1989re}, which is also related to the existence of finite dimensional matrix
representations of the deformed Heisenberg algebra \cite{Plyushchay:1997ty}. Our achievement in this section  has been to understand
the structure of the  algebra of projectors in this critical limit using the tools of symbol calculus.


\begin{figure}[ht]
\setlength{\unitlength}{.7cm}
\begin{picture}(10,10)(-5,-1)
\put(0,5){\vector(1,0){10}}
\put(5,0){\vector(0,1){10}}
\put(4.6,10.2){\large$n$}
\put(10.3,4.7){\large$m$}
\multiput(5.2,5.4)(2,0){3}{\multiput(0,0)(0,2){3}{\tiny $++$}}
\multiput(4.2,4.4)(-2,0){3}{\multiput(0,0)(0,-2){3}{\tiny $++$}}
\multiput(6.2,6.4)(2,0){2}{\multiput(0,0)(0,2){2}{\tiny $- -$}}
\multiput(3.2,3.4)(-2,0){2}{\multiput(0,0)(0,-2){2}{\tiny $- -$}}
\multiput(6.2,5.4)(2,0){2}{\multiput(0,0)(0,2){3}{\tiny $- +$}}
\multiput(3.2,4.4)(-2,0){2}{\multiput(0,0)(0,-2){3}{\tiny $- +$}}
\multiput(5.2,6.4)(2,0){3}{\multiput(0,0)(0,2){2}{\tiny $+ -$}}
\multiput(4.2,3.4)(-2,0){3}{\multiput(0,0)(0,-2){2}{\tiny $+ -$}}
\multiput(10,10)(.1,.1){10}{$\cdot$}
\multiput(0,0)(-.1,-.1){10}{$\cdot$}
\multiput(5.5,10)(1,0){5}{\multiput(0,0)(0,.1){10}{$\cdot$}}
\multiput(0.5,0)(1,0){5}{\multiput(0,0)(0,-.1){10}{$\cdot$}}
\multiput(10.3,9.4)(0,-1){5}{\multiput(0,0)(.1,0){10}{$\cdot$}}
\multiput(-.3,4.4)(0,-1){5}{\multiput(0,0)(-.1,0){10}{$\cdot$}}
\end{picture}
\caption{\small The two sets of singular quasi-projectors $\widehat T^{(-1)^m,(-1)^n}_{\l^+_{\e,m}|\e|\l^+_{\e,n}}$
for $\hat\s=+1$.
These are isomorphic to the two sub-ideals
drawn in Fig. \ref{fig3}.}
\label{fig2}
\end{figure}
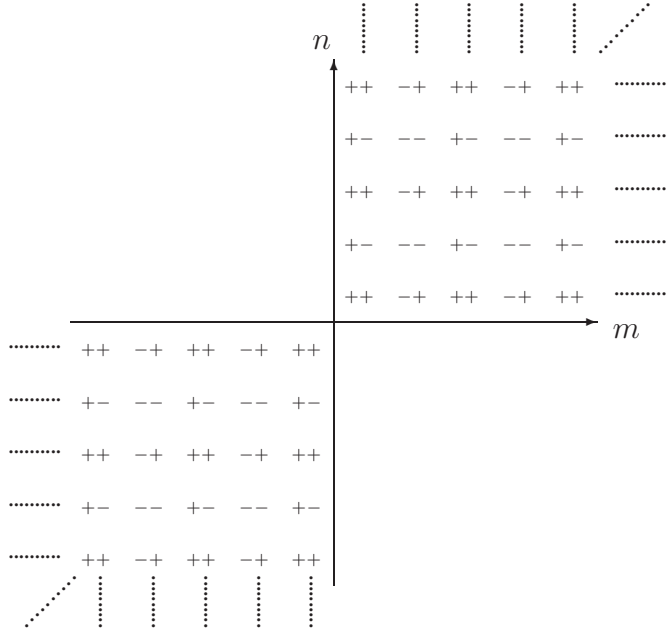


\begin{figure}[ht]
\setlength{\unitlength}{.7cm}
\begin{picture}(10,14)(-5,-3)
\put(-2,5){\vector(1,0){15}}
\put(5,-2){\vector(0,1){14.5}}
\put(13,4.5){\large$m$}
\put(4.6,12.5){\large$n$}
\multiput(5.2,5.4)(2,0){3}{\multiput(0,0)(0,2){3}{\tiny $- -$}}
\multiput(4.2,4.4)(-2,0){3}{\multiput(0,0)(0,-2){3}{\tiny $- -$}}
\multiput(6.2,6.4)(2,0){3}{\multiput(0,0)(0,2){3}{\tiny $+ +$}}
\multiput(3.2,3.4)(-2,0){3}{\multiput(0,0)(0,-2){3}{\tiny $+ +$}}
\multiput(6.2,5.4)(2,0){3}{\multiput(0,0)(0,2){3}{\tiny $+ -$}}
\multiput(3.2,4.4)(-2,0){3}{\multiput(0,0)(0,-2){3}{\tiny $+ -$}}
\multiput(5.2,6.4)(2,0){3}{\multiput(0,0)(0,2){3}{\tiny $- +$}}
\multiput(4.2,3.4)(-2,0){3}{\multiput(0,0)(0,-2){3}{\tiny $- +$}}
\multiput(5.4,10.7)(1,0){6}{\multiput(0,0)(0,.1){10}{$\cdot$}}
\multiput(-0.5,-1)(1,0){6}{\multiput(0,0)(0,-.1){10}{$\cdot$}}
\multiput(11.1,10.4)(0,-1){6}{\multiput(0,0)(.1,0){10}{$\cdot$}}
\multiput(-1.3,4.4)(0,-1){6}{\multiput(0,0)(-.1,0){10}{$\cdot$}}
\multiput(11.0,10.6)(.1,.1){10}{$\cdot$}
\multiput(-1.,-.9)(-.1,-.1){10}{$\cdot$}
\multiput(8,5)(0,.1){30}{$\cdot$}
\multiput(5,7.9)(.1,0){30}{$\cdot$}
\multiput(2,4.8)(0,-.1){30}{$\cdot$}
\multiput(4.9,1.9)(-.1,0){30}{$\cdot$}
\multiput(5,5)(0,.1){30}{$\cdot$}
\multiput(5,5)(.1,0){30}{$\cdot$}
\multiput(4.85,4.8)(0,-.1){30}{$\cdot$}
\multiput(4.9,4.8)(-.1,0){30}{$\cdot$}
\put(2,2){\line(-1,0){4}}
\put(2,2){\line(0,-1){4}}
\put(8,8){\line(1,0){4}}
\put(8,8){\line(0,1){4}}

\end{picture}
\caption{\small The decomposable structure of the set of quasi-projectors $ \widehat T^{-(-1)^m,-(-1)^n}_{\l^-_{\e,m}|\e|\l^-_{\e,n}}$
for $\hat\s=+1$ and $\ell=1$.
The coset consists of the quasi-projectors encircled by dashed lines.
Within the remaining ideal there resides two sub-ideals,
separated by solid lines, that are isomorphic to the two
sets of singular quasi-projectors depicted in Fig. \ref{fig2}.}
\label{fig3}
\end{figure}
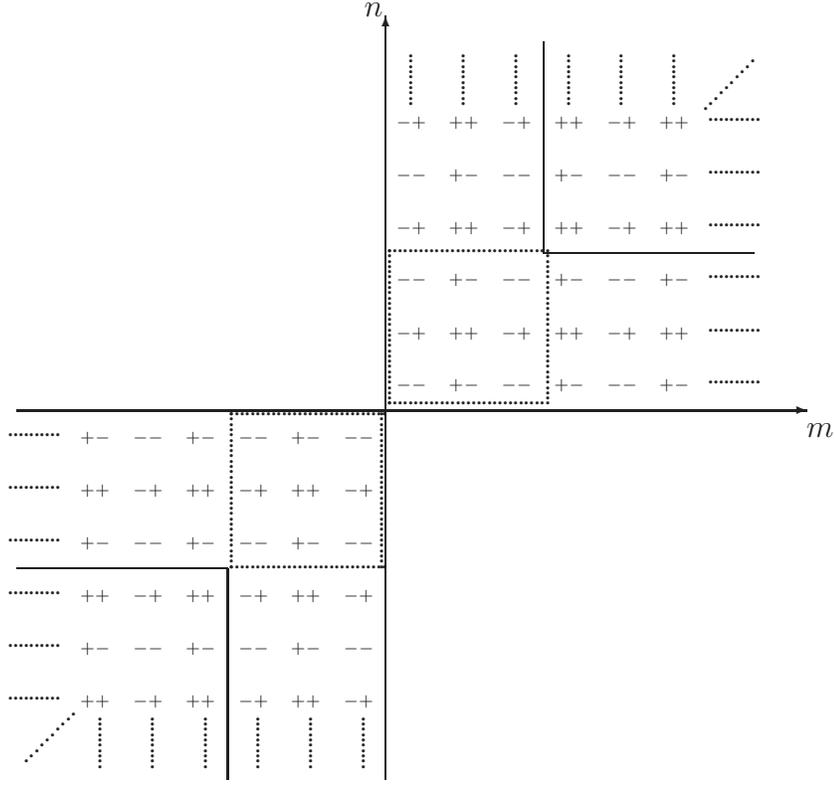

\subsection{Example: $\nu=-3$}

If $\nu=-3$, that is, if $\ell=1$ and $\hat\s=-1$, then
the basic contraction formulae read:
\bea w\star \left[w^m\right]^{+,+}&=&\left[ w^{m+1}-\frac{m^2(m+2)(m-2)}{(2m+1)(2m-1)} w^{m-1}
\right]^{+,+}\ ,\\[5pt]
 w\star \left[w^m\right]^{-,-}&=&\left[ w^{m+1}-\frac{m^2(m+1)(m-1)}{(2m+1)(2m-1)} w^{m-1}
\right]^{-,-}\ ,\\[5pt]
 a^+\star \left[w^m\right]^{+,+}\star a^-&=&\left[
w^{m+1}-(m+2) w^m+ \frac{m^2(m+1)(m+2)}{(2m+1)(2m-1)} w^{m-1}
\right]^{-,-}\ ,\\[5pt]
 a^+\star \left[w^m\right]^{-,-}\star a^-&=&\left[
w^{m+1}-(m-1) w^m+ \frac{m^2(m-1)(m-2)}{(2m+1)(2m-1)} w^{m-1}
\right]^{+,+}\ .\eea
The generalized lowest weight state projector
\be \widehat T^+_+ = \left[{}_1 F_1 \left(\frac32;3;-2w\right)\right]^{+,+}\ ,\qquad a^-\star \widehat T^+_+=0\ ,
(w-1)\star \widehat T^+_+=0\ .\ee
The descendant quasi-projectors within the triplet representation are given by
\bea \widehat T^{-,-}_{0|+|0}&=&a^+\star \widehat T^+_+\star a^-=-4 \left[{}_1 F_1\left(
\frac32;2;-2w\right)\right]^{-,-}\ ,\\[5pt]
\widehat T^{+,+}_{1|+|1}&=&a^+\star T^{-,-}_{0|+|0}\star a^-=-4\left[
\left(1-2w\frac{d}{dw}-w^2\frac{d^2}{dw^2}\right)\widehat T^+_+\right]^{+,+}\ ,\eea
which indeed have $w$ eigenvalues in accordance
with the notation.
Descending once more, we find
\be \widehat T^{-,-}_{2|+|2}=a^+\star \widehat T^{+,+}_{1|+|1}\star a^{-}=
32 \widehat T^{-}_+\ ,\ee
where the singular quasi-projector
\be \widehat T^{-}_+=\left[w\, {}_1 F_1\left(\frac52;2;-2w\right)\right]^{-,-}\ .\ee
One can also check the agreement between
computing the supertraces of the above equations
either by direct evaluation or by using the
graded cyclicity followed by $a^-\star a^+=w+\tfrac12(1+\nu k)$.

\subsection{The hypercritical cases: $\nu=\pm1$}

In the hypercritical case we have
\be \widehat T^{\hat\s}_\e=\lim_{\nu\rightarrow\hat\s}
\left[{}_1 F_1\left(\frac32;\frac{3-\hat\s\nu}2;-2\e w\right)\right]^{\hat\s,\hat\s}=
\left[{}_1 F_1\left(\frac32;1;-2\e w\right)\right]^{\hat\s,\hat\s}\ ,\ee
\be \widehat T^{-\hat\s}_\e=\lim_{\nu\rightarrow\hat\s}
\left[{}_1 F_1\left(\frac32;\frac{3+\hat\s\nu}2;-2\e w\right)\right]^{\hat\s,\hat\s}=
\left[{}_1 F_1\left(\frac32;2;-2\e w\right)\right]^{\hat\s,\hat\s}\ ,\ee
with $w$ eigenvalues
\be w\star \widehat T^{-\hat\s}_\e=0\ ,\qquad (w-\e)\star
\widehat T^{\hat\s}_\e=0\ ,\ee
as can be see explicitly using
\bea w\star \left[w^m\right]^{\hat\s,\hat\s}&=&\left[w^{m+1}-\frac{m^4}{(2m+1)(2m-1)}
w^{m-1}\right]^{\hat\s,\hat\s}\ ,\\[5pt]
w\star \left[w^m\right]^{-\hat\s,-\hat\s}&=&\left[w^{m+1}-\frac{m^2(m^2-1)}{(2m+1)(2m-1)}
w^{m-1}\right]^{-\hat\s,-\hat\s}\ .\eea
Using
\be a^{+\e}\star \left[w^m\right]^{-\hat\s,-\hat\s}\star a^{-\e}
 = \left[w^{m+1}-\e(m+1)w^m+ \frac{m^3(m+1)}{(2m+1)(2m-1)}
 w^{m-1}\right]^{\hat\s,\hat\s}\ ,\ee
one can show that
\be a^{+\e} \star \widehat T^{-\hat\s}_\e\star a^{-\e}=
-2\e \widehat T^{\hat\s}_\e\ ,\ee
which is indeed consistent with supertraceability.
Finally, we have
\be \widehat P^{-\hat\s}_\e=\widehat T^{-\hat\s}_\e\ ,\qquad
\widehat T^{\hat\s}_\e\star \widehat T^{\hat\s}_\e=0\ ,\ee
that is, the element $\widehat T^{\hat\s}_\e$ is a singular
quasi-projector while $\widehat T^{-\hat\s}_\e$ is normalizable.

\section{Final specification of the model}\label{sec:Real}

In this section, we provide the final specifications
of the model, in the form of reality conditions and
further projections of odd spins.
We then compute the resulting relation between the
higher spin gravitational and internal gauge couplings.

\subsection{Real forms}


A real form of the model can be obtained
by imposing \cite{Boulanger:2013naa}
\be \mathbb A^\dagger=-\mathbb C\star \mathbb A\star \mathbb C\ ,\qquad
\mathbb C=\left[\begin{array}{cc} 1& 0\\ 0& C\star \Pi^-\end{array}
\right]\ ,\ee
where the generalized charge conjugation matrix is given by
\eqref{cmatrix} and
\be \gamma^\dagger=\gamma\ ,\qquad \xi^\dagger=\xi\ ,\ee
and we assume the standard action of $\dagger$ on ${\rm Mat}_2$.
Thus, in terms of the separate master fields one has
\be W^\dagger=-W\ ,\qquad U^\dagger=-C\star U\star C\ ,\ee
\be \psi^\dagger=-C\star \bar\psi\ ,\qquad
\bar\psi^\dagger=-\psi\star C\ .\ee
As for the connections, this implies that
\be W\in \overline{hs}^{+}(\nu)_{(+)}\oplus
\overline{hs}^+(\nu)_{(-)}\ ,
\qquad  U\in \bigoplus_{\e,\s} \left[u(\nu;\s,\e)_{(+)}\oplus
u(\nu;\s,\e)_{(-)}\right] \ ,\ee
where $\overline{hs}^{+}(\nu)_{(\pm)}
=\Pi^+\star \overline{hs}(\nu)\star \Pi^+\star \frac12(1\pm \gamma)$
with $\overline{hs}(\nu)$ being the
power-series extension of the real form
\be hs(\nu)=\left\{f\in Aq(2;\nu)\mid f^\dagger=-f\right\}
\ ,\ee
and $u(\nu;\s,\e)_{(\pm)}=u(\nu \l;\s,\e)\star \frac12(1\pm \gamma)$
where
\be u(\nu;\s,\e)=\left\{f=\sum_{m,n=}^\infty f_n{}^m P(\nu;\s,\e)_m{}^n\;\mid\; (f_m{}^n)^\dagger=-f_n{}^n\right\}\ ,\ee
using the simplified notation
\be P(\nu;\s,\e)_m{}^n=\left(P^{-,-}_{\e}\right)_{\l^\s_{\e,2m+(1+\s)/2}}
{}^{\l^\s_{\e,2n+(1+\s)/2}}\ ,\ee
for the generalized projectors defined in \eqref{Proj}.
In this notation, we recall that
\bea P(\nu;\s,\e)_m{}^n\star P(\nu;\s,\e)_{m'}{}^{n'}&=&
\delta_{m'}^{n} P(\nu;\s,\e)_m{}^n\ ,\\[5pt]
\left(P(\nu;\s,\e)_m{}^n\right)^\dagger&=&C\star P(\nu;\s,\e)_n{}^m\star C\ ,\eea
and for non-critical $\nu$ we have
\be {\rm STr}_{\overline{Aq}(2;\nu)}
P(\nu;\s,\e)_m{}^n = -\sigma \delta_m^n \ .\ee
As for the fractional spin fields, expanding the $\xi$ dependence as in \eqref{fmodel} and \eqref{bmodel}, we have
\be \overline\Theta^\dagger=-\Theta\star C\ ,\qquad
\overline\Sigma^\dagger=-C\star \Sigma\ .\ee
The equations of motion thus take the form
\bea \mbox{${\cal A}_+$ model:}&& F^W-\Theta\star C\star\Theta^\dagger=0\ ,\quad
F^U-C\star\Theta^\dagger\star \Theta=0\ ,\quad D\Theta=0\ ,\label{eomA+}\\[5pt]
\mbox{${\cal A}_-$ model:}&& F^W-\Sigma\star C\star\Sigma^\dagger=0\ ,\quad
F^U-C\star\Sigma^\dagger\star \Sigma=0\ ,\quad D\Sigma=0\ .\label{eomA-}\eea
where $F^W=dW+W\star W$ and $F^U=dU+U\star U$.

\subsection{$\tau$ projections}


One way of truncating further the resulting model is
to remove components from $(\psi,\overline\psi;U)$
that have distinct eigenvalues of $w$; for example, one
may restrict $U$ to $u(\nu;+,+)$ and $(\psi,\bar\psi)$
accordingly, and possibly proceed by further level truncations.
However, taking into account the nature of the critical limits,
and the fact that truncations by algebra (anti-)automorphism
would be more natural from the point-of-view of an underlying
topological open string, it is more natural to seek to truncate
the model by extending the map \eqref{taumap} to the fractional spin
algebra.
This can be achieved by letting $\tau$ act on ${\rm Mat}_2$ as matrix
transposition and taking\footnote{
The Clifford algebra ${\rm Cliff}_N(\xi^i$) with $N$ fermionic generators
$\xi^i$ obeying $\{\xi^i,\xi^j\}_\star=2\delta^{ij}$
has a graded anti-automorphism defined by $\tau(f\star g)=
(-1)^{\e_s(f)\e_s(g)}\tau(g)\star\tau(f)$ and $\tau(\xi^i)=i\xi^i$
where $\e_s$ denotes the Grassmann statistics and $\e_s(\xi^i)=1$.}
\be \tau(\gamma)=\gamma\ ,\qquad \tau(\xi)=i\xi\ ,\label{tauext}\ee
from which it follows that
\be \tau(\mathbb X\star \mathbb X')=
\tau(\mathbb X')\star \tau(\mathbb X)\ .\label{taufs}\ee
Thus, in the ${\cal A}_+$ model we may impose the
following $\tau$ projection\footnote{
The ${\cal A}_-$ model, in which $\tau^2(\psi,\overline\psi)=-(\psi,\overline\psi)$,
cannot be projected using the basic $\tau$ map defined
by \eqref{taumap}, \eqref{tauext} and matrix transposition
in ${\rm Mat}_2$.}
\be \mbox{${\cal A}_+$ model:}\ \tau(\mathbb A)=-\mathbb A\ ,\ee
or, equivalently, in terms of the separate master fields,
\be \tau(W)=-W\ ,\qquad \tau(U)=-U\ ,\qquad \tau(\Theta)=-i\overline\Theta\ ,\qquad
\tau(\overline\Theta)=-i\Theta\ ,\ee
which are indeed consistent with $\tau^2=\pi_q\pi_\xi$ as well as
the equations of motion \eqref{eomA+}.
The $\tau$ projection removes the fields in $W$ with
odd spin and relates the fields in $U$ arising
from the gauging of $u(\nu;\s,\e)$ and
$u(\nu;\s,-\e)$, respectively, which
we denote as
\be W\in \overline{hs}^{+}_0(\nu)_{(+)}\oplus
\overline{hs}^+_0(\nu)_{(-)}\ ,\qquad \overline{hs}^{+}_0(\nu)
=\left\{f\in \overline{hs}^{+}(\nu)\;\mid\;
\tau(f)=-f\right\}\ ,\ee
\be U\in \sum_{\s} \left[u(\nu;\s)_{(+)}\oplus
u(\nu;\s)_{(-)}\right]\ ,\qquad u(\nu;\s)=
u(\nu;\s,+)-\tau(u(\nu;\s,+))\ ,\ee
using the fact that $\tau(u(\nu;\s,\e))=u(\nu;\s,-\e)$
and $\tau^2(u(\nu;\s,\e))=u(\nu;\s,\e)$.
%

\subsection{Final form of action and couplings}

Turning to the gauge couplings, up to boundary terms
that we will not specify here, the canonically normalized
higher spin gravitational and internal parts of the action
are defined by
\bea S_{\rm hs}[W] &=&\frac{k_{\rm hs}}{2\pi}\int_{M_3}
{\rm Tr}_{hs^+(\nu)} \left[\tfrac12 \,W\star dW+\tfrac13 \,W^{\star 3}\right]\ ,\\[5pt]
S_{\rm int}[U]&=&\frac{k_{\rm int}}{2\pi}\int_{M_3}
{\rm Tr}_{u(\nu)} \left[\tfrac12 \, U\star dU+\tfrac13 \, U^{\star 3}\right]
\ ,\eea
using trace operations normalized as follows:
\be {\rm Tr}_{hs^+(\nu)} (J_a\star J_b)=\frac12 \;\eta_{ab}\ ,\qquad
{\rm Tr}_{u(\nu)} (T_m{}^n\star T_{m'}{}^{n'})=\frac12 \;\delta_m^{n'}\delta_{m'}^n\ .\label{normTr}\ee
In the gravitational sector, the resulting expressions for Newton's constant and the cosmological constants read
\begin{equation}
G_{\rm N} = \frac{\ell_{\rm AdS}}{4k_{\rm hs}}\;,\qquad \Lambda =
-\frac{1}{\ell^2_{\rm AdS}}\;,
\label{GN}\end{equation}
while the internal level must be integer provided that the base
manifold $M_3$ is compact and orientable.
Comparing \eqref{normTr} to
\be {\rm STr}_{{Aq}(2;\nu)} \left(J_a\star J_b\right)=
\frac1{32}\,(1-\nu^2)(1-\frac\nu3)\, \eta_{ab}\ ,\ee
which follows using \eqref{Ja} and
\be Tr_{Aq(2;\nu)} \left(J_{\alpha(2)} \star J_{\beta(2)}\right)
=-\tfrac{1}{4}\,\epsilon_{\alpha\beta}\epsilon_{\alpha\beta}\;
(1-\nu^2)(1-\frac\nu 3)\ ,\ee
and
\be {\rm STr}_{Aq(2;\nu)} \left(\left(P(\nu;\s,\e)\right)_{m}{}^n \star
\left(P(\nu;\s,\e)\right)_{m'}{}^{n'} \right) =-\sigma  \delta_m^{n'}\delta_{m'}^n\ ,\ee
it follows that
\bea {\rm STr}_{Aq(2;\nu)}|_{{\rm hs}^+(\nu)}&=&\frac1{16}(1-\nu^2)(1-\frac\nu3) \,{\rm Tr}_{{\rm hs}^+(\nu)}\ ,\\[5pt]
 {\rm STr}_{Aq(2;\nu)}|_{u(\nu;\s,\e)}&=&-2\s \,{\rm Tr}_{u(\nu)}\ .\eea
Hence we obtain
\be
k_{\rm hs}=\frac{\varkappa}{16}(1-\nu^2)(1-\frac\nu 3)\ ,\qquad
 k_{\rm int}(u(\nu;\s,\e))=\mp \sigma \varkappa \ ,\label{couplings}\ee
where the $\mp$ sign is correlated to the fermion/boson model as one can see  from \eqref{action2}.

Finally, the $\tau$ projected ${\cal A}_+$ model has the following critical limits:
\bea \nu=-(2\ell+1)&:& \mathbb A\in \left[\begin{array}{cc} gl(\ell+1) & (\ell+1,\ell)\\[5pt]
(\ell,\ell+1)&gl(\ell)\end{array}\right]=gl(\ell+1|\ell)\ ,\\[5pt]
\nu=2\ell+1&:& \mathbb A\in \left[\begin{array}{cc} gl(\ell) & (\ell,\ell+1)\\[5pt]
(\ell+1,\ell)&gl(\ell+1)\end{array}\right]=gl(\ell|\ell+1)\ .\eea
Indeed, the gravitational level vanishes, \emph{i.e.}
Newton's constant diverges, in all
cases when the $W$ field belongs to either $gl(0)$ or $gl(1)$.

\section{Conclusion}\label{sec:Conclusion}

In summary, we have taken the first steps towards equipping 
the fractional spin algebra introduced in \cite{Boulanger:2013naa}
with a trace operation suitable for the construction
of a Chern-Simons model that unifies higher spin gravity,
internal gauge fields and fractional spin fields.
The model is reminiscent to ordinary Chern--Simons 
gauged supergraviy \cite{Achucarro:1987vz} and its chirally
asymmetric versions \cite{Giacomini:2006dr}.
Indeed, as the fractional spin parameter $\nu$ varies,
the model interpolates $gl(\ell+1|\ell)$ and $gl(\ell|\ell+1)$
models, which arise for the critical values $\nu=-2\ell-1$
and $\nu=2\ell+1$, respectively.
In particular, we have used the star product formalism to
obtain the relation between the gravitational and
internal gauge couplings.

In a more abstract sense, the unification of tensorial
and fractional spin fields is possible owing to the existence
of different realizations of $sl(2,\mathbb{R})$ in the enveloping algebra of
the Wigner-deformed Heisenberg algebra \eqref{dha} arising upon choosing
different bases.
The present model incorporates two such domains of $sl(2,\mathbb{R})$,
namely the class of polynomial elements associated with Lorentz
tensorial fields, and a class of Gaussian elements associated with
fractional spin and internal gauge fields.
Along the construction, we have made several choices,
whereas a full classification of all possible models
would require a more thorough study of the action of
$sl(2,\mathbb{R})$ on its enveloping algebra and corresponding
supertrace operations.
To our best understanding, this remains an open problem,
at least in the context of physical model building.

To make further progress, and in particular to
enrich the $sl(2,\mathbb R)$ modules by additional sectors,
it would be desirable to use the star product and
supertrace operations \emph{sensu amplo}.
To this end, one may use the method employed
in this paper, namely to extend the polynomial enveloping
algebra by non-polynomial elements corresponding
to endomorphisms in lowest-energy spaces spanned 
by eigenstates of a Hamiltonian belonging to 
the enveloping algebra of $sl(2,\mathbb R)$.
The salient feature of the resulting extended
algebra, namely its associativity and traceability, 
then follow from the existence of a ground state 
projector given by a traceable, \emph{i.e.}
real-analytic, element.
In the present paper, the latter condition amounts
to the finiteness of the normalization coefficient
in \eqref{key}, which we checked to the first 2 orders 
in $w$ and which we hope to demonstrate in
its entirety in a forthcoming publication within 
the context of a more convenient
realization of the star product.

As for holographic duality, a boundary analysis, possibly along the lines of \cite{Henneaux:2010xg, Campoleoni:2010zq,Gaberdiel:2010ar},
will reveal whether there exist conditions consistent with the
standard ones for internal gauge fields and higher spin fields.
If so, we expect there to exist a map from boundary states to
conformal current algebras,
including stress tensors, Kac--Moody currents
and intertwining currents \cite{Jatkar:1990re,Moore:1991ks}.
In particular, it would be interesting to exhibit
the boundary states generated by the fractional spin
fields, and also to understand whether and how the level of
the internal gauge connection, and hence Newton's constant,
could be quantized.
The holographically dual description may also shed light on the subtle
fact that the ideals that arise in critical limits disappear from the
action, which amounts to limits of infinite coupling.

The results here presented can be extended to formulations of non-topological theories, namely of the type \cite{Prokushkin:1998bq}, or by introducing massive anyons (e.g. of \cite{Plyushchay:1994re,Jackiw:1990ka,Cortes:1992fa})  in a fractional higher spin gravity background, which we expect to present elsewhere.

\section*{Acknowledgments}

We thank Thomas Basile, Xavier Bekaert, Fabien Buisseret,
Slava Didenko, Dileep Jatkar, Antal Jevicki and Tomas Prochazka for discussions.
For are especially grateful to Thomas Basile for collaboration 
in proving the relation \eqref{3.40}. 
N.B. is supported by the ARC contract N$^{\rm o}$ AUWB-2010-10/15-UMONS-1.
P.S. is supported by Fondecyt Regular grant N$^{\rm o}$ 1140296 and Conicyt grant DPI 20140115.
M.V. acknowledges the hospitality of UNAB (Santiago) and CECs (Valdivia).

\begin{appendix}

\section{Conventions}\label{app:A}

\noindent We essentially follow the conventions and notation of \cite{Achucarro:1989gm}.
We work with metric $\eta_{ab}$ and epsilon symbols $\varepsilon_{abc}$
and $\varepsilon^{abc}$ obeying
\be \eta_{ab}={\rm diag}(-++)_{ab}\ ,\qquad
\varepsilon^{abc}\varepsilon_{a'b'c'}=-3!\delta^{[a}_{a'}\delta^{\,b}_{\,b'} \delta^{c]}_{c'}\ ,\ee
and generators
\be L_{ab}=-\varepsilon_{abc} J^c\ ,\qquad J_a=\frac12 \varepsilon_{abc} L^{bc}\ee
of the Lorentz group obeying
\begin{equation}
[L_{ab},L^{cd}]=4i\delta^{[c}_{[b} L_{a]}{}^{d]}\ ,\qquad
[J^a , J^b] = - i\, \varepsilon^{abc} J_c\;.
\label{alg0}
\end{equation}
We use real van der Waerden symbols
\be (\tau^a)_{\a\b}=(\tau^a)_{\b\a}=\left((\tau^a)_{\a\b}\right)^\dagger\ ,\qquad
\e_{\a\b}=-\e_{\b\a}=\left(\e_{\a\b}\right)^\dagger\ee
obeying
\be (\tau^a)_\a{}^\b(\tau^b)_{\b}{}^{\g}=\eta^{ab}\d_{\a}^{\g}+
\varepsilon^{abc}(\tau_c)_{\a}{}^{\g}\ ,\qquad
\epsilon^{\a\b}\e_{\g\d}=2\delta^{\a}_{[\g}\delta^\b_{\d]}\ ,\ee
using the convention $q^\a=\e^{\a\b}q_\b$.
A convenient realization in terms of the Pauli matrices
is
\be (\tau^a)_{\a\b}=(-1,\sigma^1,\sigma^3)_{\a\b}\ ,\qquad
(\tau^a)_{\a}{}^{\b}=(-i\s^2,-\sigma^3,\sigma^1)_{\a\b}\ ,\label{taumatrices}\ee
\be \e_{\a\b}=(-i\s^2)_{\a\b}\ ,
\qquad \e^{\a\b}=(-i\s^2)^{\a\b}\ ,\qquad \varepsilon^{012}=1\ .\ee
In this realization, the operator $w=2J_0=\frac14( q^1\star q^1+q^2\star q^2)$,
\emph{i.e.} the Hamiltonian of the deformed harmonic oscillator with
deformed momentum $\frac1{\sqrt{2}}q^1$ and coordinate $\frac1{\sqrt{2}} q^2$.

\end{appendix}

\providecommand{\href}[2]{#2}\begingroup\raggedright\endgroup

\end{document}